\documentclass[prd,showpacs,nofootinbib,amsmath,amssymb,floatfix,superscriptaddress,showkeys]{revtex4}
\usepackage{multirow}
\usepackage{graphicx}
\usepackage{epstopdf}
\usepackage{dcolumn}
\usepackage{bm}
\usepackage{threeparttable}
\usepackage{subfigure}
\usepackage{color,txfonts}
\usepackage{ulem}
\definecolor{blue0}{rgb}{0,0,0.6}
\usepackage[colorlinks,linkcolor=blue0,anchorcolor=blue0,citecolor=blue0,urlcolor=blue0]{hyperref}

\newcommand{\jcap}{J. Cosmol. Astropart. Phys.}

\newcommand{\aap}{Astron. Astrophys}

\newcommand{\beq}{\begin{equation}}
\newcommand{\eeq}{\end{equation}}
\newcommand{\beqa}{\begin{eqnarray}}
\newcommand{\eeqa}{\end{eqnarray}}

\begin{document}

\title{Search for gamma-ray emission from the 12 nearby dwarf spheroidal galaxies with 12 years of Fermi-LAT data}

\author{Shang Li}
\affiliation{School of Physics and Optoelectronics Engineering, Anhui University, Hefei 230601}
\author{Yun-Feng Liang}
\email[]{Corresponding author: liang-yf@foxmail.com}
\affiliation{Laboratory for Relativistic Astrophysics, Department of Physics, Guangxi University, Nanning 530004, China}

\author{Yi-Zhong Fan}
\email[]{Corresponding author: yzfan@pmo.ac.cn}
\affiliation{Key Laboratory of Dark Matter and Space Astronomy, Purple Mountain Observatory, Chinese Academy of Sciences, Nanjing 210008, China}
\affiliation{School of Astronomy and Space Science, University of Science and Technology of China, Hefei, 230026, China}

\date{\today}

\begin{abstract}

Previously, we have shown in Li et al. (2018) that very weak $\gamma$-ray excesses ($\sim 2\sigma$) appear in some Milky Way dwarf spheroidal galaxies (dSphs, including candidates) and the combination analysis of 12 nearby dSphs yields a local significance of $>4\sigma$.
In this work, we adopt a longer data set (i.e., the 12 years of Fermi-LAT  data), the latest Fermi-LAT software as well as background models to update the searches of $\gamma$-ray emission from these sources. Very weak $\gamma$-ray excesses ($>$ 2 $\sigma$) are found in the directions of three dSphs, including Reticulum II, Bootes II and Willman 1. In the direction of Reticulum II, the peak TS value of the excess reaches $\sim$11. However, different from the previous analysis with the 9 years of Fermi-LAT data, now the location of the gamma-ray emission is significantly away from the center of Reticulum II because of the enhancement of ``offset" $\gamma$-rays above 10 GeV since 2017. The detected weak excess is likely due to the contamination of an astrophysical $\gamma$-ray source with a TS value of $\sim 22$, irrelevant to the dark matter inside Reticulum II. The possible excesses in the directions of Bootes II and Willman 1 are weaker with lower peak TS values ($\sim 7$). If interpreted as annihilation of dark matter particles into $\tau^+\tau^-$,  the dark mass mass of $m_\chi \sim 14$ GeV and $\sim 80$ GeV are found for Bootes II and Willman 1, respectively. Much more data are needed to clarify whether these two potential signals are real and then reveal their origins.\end{abstract}

\maketitle
\section{Introduction}

Many astrophysical observations suggest that there is a large amount of dark matter (DM) in the universe. According to the latest observation results, non-baryonic cold dark matter (DM) constitutes $\sim$ 84\% of the matter density of the Universe \cite{Ade:2015xua}. The nature of DM particles is still unknown and the most promising DM candidates are weakly interacting massive particles (WIMPs). WIMPs can annihilate or decay into Standard Model (SM) particles and finally produce GeV-TeV $\gamma$-rays or cosmic rays \cite{Jungman:1995df, Bertone:2004pz, Hooper:2007qk, Feng:2010gw,PhysRevD.69.123501}. The primary aim of the dark matter indirect detection is to identify these products from a dark matter origin through astronomical experiments such as the Fermi Large Area Telescope (Fermi-LAT \cite{atwood09lat}) and the Dark Matter Particle Explorer (DAMPE \cite{DAMPE:2017,Chang:2017n}).

The Milky Way dwarf spheroidal (dSph) galaxy is considered to be one of the most promising targets for indirect detection of DM. The reasons include that they are close to us ($<100\,{\rm kpc}$ for many of them) and the kinematic observations show that they are DM-dominated systems. In addition, benefit from the lack of astrophysical $\gamma$-ray production mechanisms \cite{Lake:1990du,Baltz:2004bb,Strigari:2013iaa}, the diffuse $\gamma$-ray background for searching for DM signal with dSphs is very low. For a long time, no potential evidence of DM signal was found in the direction of dSphs and based on the non-detection people have set very strong constraints on the mass $m_\chi$ and the annihilation cross section $\left<\sigma v\right>$ of the particle DM \cite{fermi11dsph,GeringerSameth:2011iw,2012PhRvD86b3528C,tsai13dsph,fermi14dsph,zhao2016ds,gs15dsph,fermi15dsph}. Especially, the stacked analysis of 15 dSphs with Fermi-LAT Pass 8 data excludes the thermal DM particle of the mass $<$ 100 GeV \cite{fermi15dsph}.

\begin{center}
\begin{table*}[!t]
\caption{The information and results of the 12 dSphs.}
\begin{tabular}{lcccccccccc}
\hline
\hline
 Name &  $(l,b)$ & ${\rm Distance}$ & $\log_{10}{{(J)}^{a}}$& $\log_{10}{({\rm Est.}\;J)^{b}}$ && ${\rm TS}^{\rm PL}$& & ${\rm TS}_{\rm peak}^{b\bar{b}}$ & & ${\rm TS}_{\rm peak}^{\tau^{+}\tau^{-}}$ \\
     & [deg] &[kpc]&[$\log_{10}{\rm (GeV^{2}cm^{-5})}$] & [$\log_{10}{\rm (GeV^{2}cm^{-5})}$] & &  & & \\
\hline
Bootes II      &(353.69, 68.87) & $42$ &$-$& 18.9  && 3.7&&    5.4  & &   6.4  \\
Bootes III     &(35.41, 75.35) & $47$ &$-$& 18.8   & &   0.0&& 2.8  & &   0.0    \\
Coma Berenices &(241.89, 83.61) & $44$ &19.0$\pm0.4$& 18.8  & & 0.4&&   0.6  & &   0.7  \\
Draco II       &(98.29, 42.88) & $24$ &$-$& 19.3   & &   0.0&& 0.5  & &   0.4  \\
Cetus II       &(156.47, -78.53) & $30$ &$-$& 19.1 & &    0.8&&3.6  & &   3.2  \\
Reticulum II   &(266.30, -49.74) & $32$ &18.9$\pm0.6$& 19.1 & &  11.0&& 10.9  & &   10.8\\
Segue 1        &(220.48, 50.43) & $23$ &19.4$\pm0.3$& 19.4  & &  0.0&&  0.0 & &   0.0  \\
Triangulum II  &(140.90, -23.82) & $30$ &$-$& 19.1 & &  0.0&&  0.0  & &   0.0  \\
Tucana III     &(315.38, -56.18) & $25$ &$-$& 19.3 & &   0.5&& 2.1  & &   2.4  \\
Tucana IV      &(313.29, -55.29) & $48$ &$-$& 18.7 & &    0.0&&0.9  & &   1.3  \\
Ursa Major II  &(152.46, 37.44) & $32$ &19.4$\pm0.4$& 19.1  & &  0.0&&  1.3  & &   0.0  \\
Willman 1      &(158.58, 56.78) & $38$ & $-$&18.9  & &   7.5&& 7.6  & &   7.3  \\
\hline
\end{tabular}
\begin{tablenotes}
\item $^{\rm a}$ J-factors derived through stellar kinematics. For Reticulum II, it is taken from \cite{Simon:2015fdw}, others are from \cite{Geringer-Sameth:2014yza}.
\item $^{\rm b}$ J-factors estimated with the empirical relation $J(d)\approx 10^{18.1\pm 0.1}(d/100~{\rm kpc})^{-2}$ \cite{Drlica-Wagner:2015xua}, where $d$ is the distance.
\end{tablenotes}
\label{tab:12dsph}
\end{table*}
\end{center}

N-body cosmological simulations suggested that there are many dSphs in the Milky Way halo. In the past few years, more than 20 new dSphs and candidates were found by several newly launched optical imaging surveys \cite{des15y1, des15y2, laevens15tri2, laevens15_3, kim15pegasus3, Daisuke(2016), Daisuke(2017), Drlica-Wagner1(2016), Torrealba(2016)}, making the total number of discovered dSphs/candidates is over 50. Many groups have searched for the $\gamma$-ray emission from these newly discovered dSphs\footnote{Hereafter we use the term dSph to express both identified dSphs and candidates for brevity.} with Fermi-LAT data \cite{gs15ret2, hooper15ret2,Drlica-Wagner:2015xua,li16dsph,fermi2016dsph,liang2016dl}. Although no significant signals were robustly found, very weak $\gamma$-ray signals were reported in the directions of Reticulum II \cite{gs15ret2, hooper15ret2,Drlica-Wagner:2015xua, fermi2016dsph,2018PhRvD..97l2001L}, Tucana III \cite{li16dsph, fermi2016dsph} and Tucana II \cite{Bhattacharjee:2018xem}. More intriguingly,  these tentative emission spectra resemble the Galactic GeV excess reported in \cite{Hooper:2010mq, Gordon:2013vta, Hooper:2013rwa, Daylan:2014rsa, Zhou:2014lva, Calore:2014xka, Huang:2015rlu, fermi17GCE,ABDUGHANI2021}.

In a previous work (\cite{2018PhRvD..97l2001L}, hereafter L18), we have studied the $\gamma$-ray emission of the 12 nearest dSphs at distances $\leq$ 50 kpc and found out that there are weak $\gamma$-ray excesses in several sources. In particular, the local significance of the $\gamma$-ray signal in Reticulum II is $>$ 3$\sigma$ and the significance is increasing with time (i.e., having a temporal behavior like a true steady source). Moreover, the combination analysis of the 12 sources yields a local test statistic (TS, see Eq.~(\ref{eq:ts}) for its definition) value of $\sim18.4$ of the tentative $\gamma$-ray emission. From then on, more data have been accumulated and the new analysis software and background templates of Fermi-LAT are released. If the weak excess reported in L18 is a real signal (no matter astrophysical or DM origin), the local significance for the combined analysis is expected to increasing to the value of TS $\sim25$ with the observation of 3 more years. Therefore, in this work we use the latest Fermi-LAT software as well as background models to search for $\gamma$-ray emission from the 12 nearby dSphs with 12 years of Fermi-LAT Pass 8 data. We aims to examine the existing tentative signals and search for new possible signals. The sample is listed in Table \ref{tab:12dsph}.

\section{data analysis}\label{data}

We use twelve years (i.e. from 2008 October 27 to 2020 October 27) of Fermi-LAT Pass 8 data in 500 MeV to 500 GeV. In order to remove the effect of the Earth’s limb, we reject the $\gamma$ events with zenith angle greater than $100^{\circ}$. Meanwhile, the quality-filter cuts ({\tt DATA\_QUAL==1 \&\& LAT\_CONFIG==1}) are applied to ensure the data can be used for scientific analysis. We take a $5^{\circ}$ region of interest (ROI) for each target. The latest version of {\tt Fermitools} is employed to analyze the Fermi-LAT data. The script {\tt make4FGLxml.py}\footnote{\url{https://fermi.gsfc.nasa.gov/ssc/data/analysis/user/python3/make4FGLxml.py}} is used to generate the background models, with all 4FGL-DR2\footnote{\url{https://fermi.gsfc.nasa.gov/ssc/data/access/lat/10yr_catalog/}} sources within 10$^{\circ}$ of the center of each target and the latest diffuse models (i.e {\tt gll\_iem\_v07.fits} and {\tt iso\_P8R3\_SOURCE\_V3\_v1.txt}) included. 
Due to the lack of information about the spatial expansion of the dark matter halos in the twelve dSphs, we model them as point-like sources.

Firstly, a standard unbinned likelihood analysis\footnote{\url{https://fermi.gsfc.nasa.gov/ssc/data/analysis/scitools/likelihood_tutorial.html}} is applied to get the best-fit parameters for the background sources. During the likelihood analysis, the parameters of all the 4FGL-DR2 sources within ROI, as well as the normalizations of the two diffuse backgrounds are set free. 
Secondly, we apply a likelihood profile method to derive the TS value and flux upper limit of the potential $\gamma$-ray emission for each target. 
We divided the whole data set in the energy range of 500 MeV - 500 GeV into 24 logarithmically-spaced energy bins. For each energy bin $k$, we derive the relation $L_k(f_k)$ between the likelihood $L_k$ and the target's flux $f_k$. A power-law spectral model ($dN/dE \propto E^{-\Gamma}$) with $\Gamma$=2 \cite{fermi15dsph} is used to model the putative dSph source. In order to get better sensitivity, we also apply an unbinned likelihood analysis to generate the likelihood profile. The likelihood profile can then be used to scan a series of DM masses and different annihilation channels in later analysis.  
 
A broadband likelihood function for DM models with parameter $\bm{\alpha}$ can be obtained by multiplying the bin-by-bin likelihoods together,

\begin{equation}
L(\bm{\alpha})=\prod_k L_k(f_k(\bm{\alpha})).
\label{eq1}
\end{equation}

The analysis method here is similar to that developed in \cite{fermi11dsph,tsai13dsph,fermi14dsph} and more details can be found in these articles. The test statistic (TS) is used to quantify the significance of the target sources, which  is defined as \cite{1996ApJ...461..396M}
\begin{equation}
{\rm TS}=-2\ln(L_{\rm bkg}-L_{\rm dsph}), 
\label{eq:ts}
\end{equation}
where the $L_{\rm bkg}$ and $L_{\rm dsph}$ are the best-fit likelihood values for the background-only model and the model containing a putative dsph, respectively.

\begin{figure*} 
\begin{center}
\includegraphics[width=0.45\textwidth]{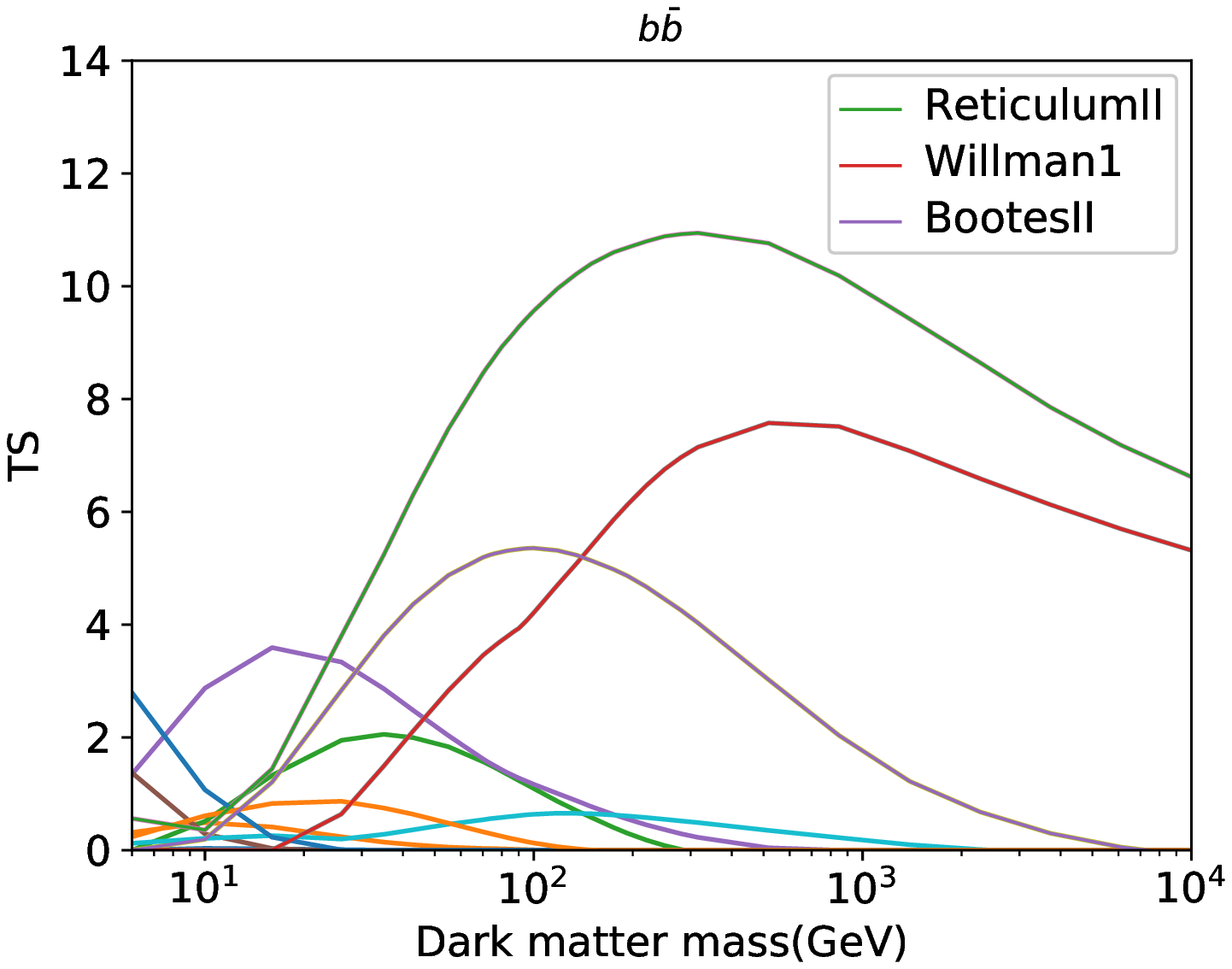}
\includegraphics[width=0.45\textwidth]{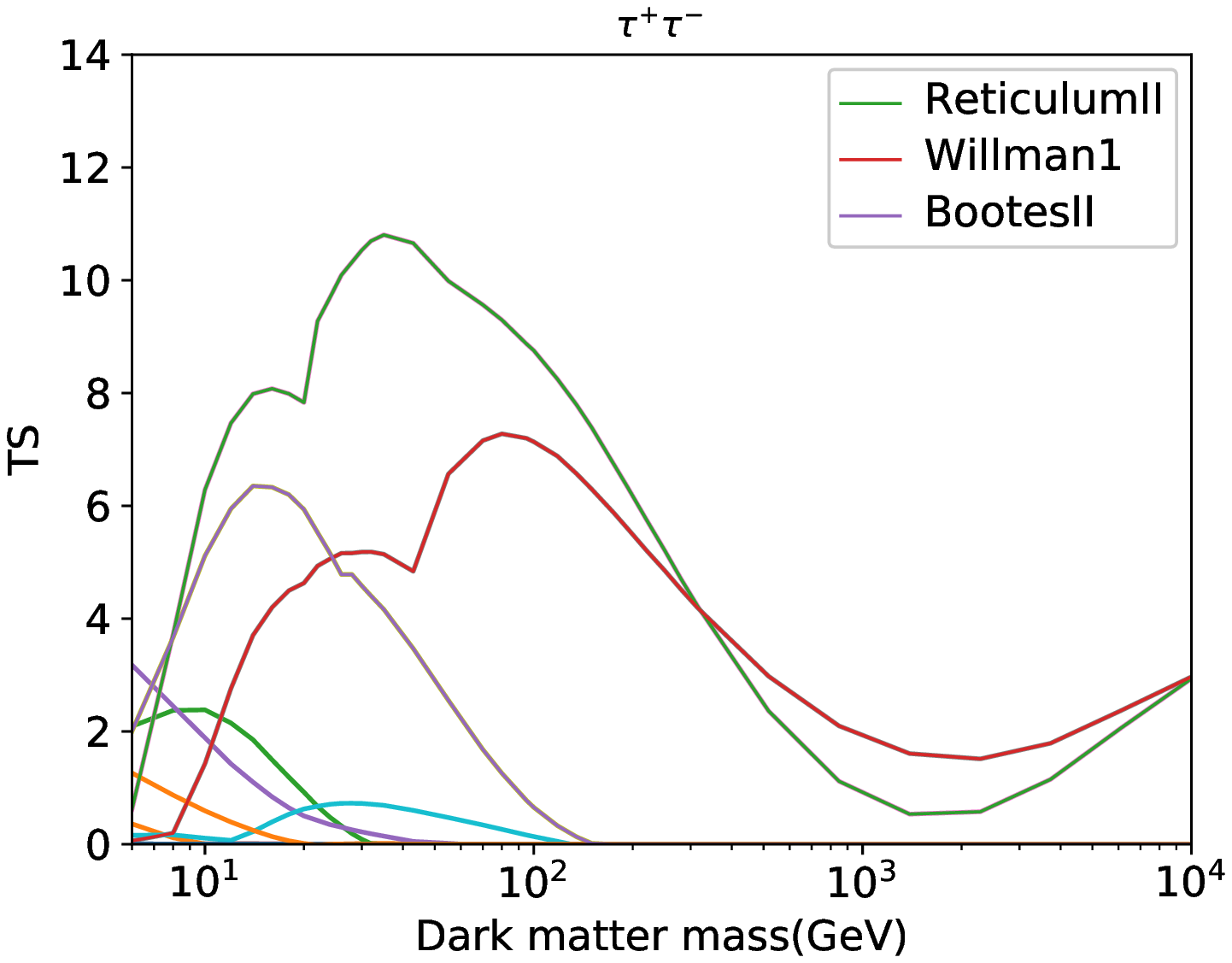}
\end{center}
\caption{ The TS values of the twelve dSphs as a function of the DM mass for two annihilation channels (i.e., $b\bar{b}$ and ${\tau^{+}\tau^{-}}$).}
\label{fig:1}
\end{figure*}


\section{Searching for dark matter emission from the 12 dSphs} 

The dSphs are good targets for DM search because the kinematic observations show that they are DM-dominated objects. The expected $\gamma$-ray flux from DM annihilation is expressed as \cite{Jungman:1995df, Bertone:2004pz, Hooper:2007qk, Feng:2010gw}
\begin{equation}
{\Phi}(E_{\gamma})={\frac{\left<{\sigma}v\right>}{8{\pi}m_{\chi}^{2}}\frac{dN_{\gamma}}{dE_{\gamma}}\times J},
\end{equation}
where ${m_{\chi}}$, ${\left<{\sigma}v\right>}$, $dN_{\gamma}/dE_{\gamma}$ are the DM particle mass, the velocity-averaged DM annihilation cross
section and the differential $\gamma$-ray spectrum per annihilation. In this paper, the DM spectra are obtained from PPP4DMID \cite{Cirelli:2010xx}. The term 
\begin{equation}
J={\int}{\rho}^{2}(r)dld{\Omega}
\end{equation}
is the line-of-sight integral of the square of the DM density (i.e., the so-called J-factor).  

In this section, we search for the $\gamma$-ray emission from each of the 12 dSphs. 
We use the likelihood profile method to scan a range of DM masses from 6 GeV to 10 TeV for two typical DM annihilation channels (i.e., ${b\bar{b}}$ and ${\tau^{+}\tau^{-}}$).
The TS values as a function of DM mass are presented in Fig.~\ref{fig:1}. Even though much more data and the latest background models are considered in our work, no significant (i.e., $\rm TS>25$) signals are found in our analyses and most of the sources result in ${\rm TS}\sim0$ (see Table \ref{tab:12dsph}). 
However, the local significances of the $\gamma$-ray emissions in the directions of three dSphs (i.e., Reticulum II, Bootes II, Willman 1) are $>$ 2$\sigma$.
The most significant excess appears in the direction of Reticulum II (${\rm TS}\sim11$, i.e. the local significance is $>$ 3.0$\sigma$), coinciding with previous results \cite{gs15ret2,hooper15ret2,Drlica-Wagner:2015xua,fermi2016dsph,2018PhRvD..97l2001L}.
The modeling of a power-law spectrum (i.e., $dN/dE \propto E^{-2}$) to the tentative $\gamma$-ray signals yields similar results.
The maximal TS values for both PL and DM models are summarized in Table \ref{tab:12dsph}. In the following, we will investigate the three sources, Reticulum II, Bootes II and Willman 1, more detailedly in Section \ref{sec:ret2}.

We also perform a combined analysis of these 12 objects. For the analysis method, we refer readers to the Sec. III B of L18 and the references therein for more details. The combined analysis can improve the sensitivity of the analysis by stacking multiple sources, and can take into account whether the signal strength in individual dSph matching the J-factor of the source. In L18, we find that the combination of the 12 nearest ($<50$ kpc) dwarf galaxies shows a possible signal of TS$\sim$18.4. We examine this tentative signal here. More specifically, if the signal is physically real (no matter astrophysical or DM origin), then adding 3 more years of data the TS value is expected to reach $\sim$25. However, in our analysis of the 12-year data, the TS value drops to $\sim$9.3 ($\sim$10.0) for $b\bar{b}$ ($\tau^+\tau^-$). The above results are based on J-factors from \cite{Simon:2015fdw} and \cite{Geringer-Sameth:2014yza}. As mentioned above, for the combined analysis, we obtain a high significance only when the tentative gamma-ray excess from every source is consistent with its expected J-factor. It is very important to accurately determine the J-factors of dSphs for the combined analysis. Therefore, we also investigate how the results will change by using the J-factor values in \cite{PhysRevD.93.103512}  or  \cite{PhysRevD.94.063521}. These works include effects like flattening, and for some sources they give J-factor values different from \cite{Simon:2015fdw} and \cite{Geringer-Sameth:2014yza}. With the \cite{PhysRevD.93.103512} and \cite{PhysRevD.94.063521} J-factors, we obtain TS values of 7.4 (8.0) and 9.0 (9.7) for the $b\bar{b}$ ($\tau^+\tau^-$) channel, which are similar to the above results. Such a result is likely not supportive of the DM origin. The excess, if not due to the statistical fluctuations, should be contributed by variable astrophysical sources.

\begin{figure}[t]
\begin{center}
\includegraphics[width=0.5\textwidth]{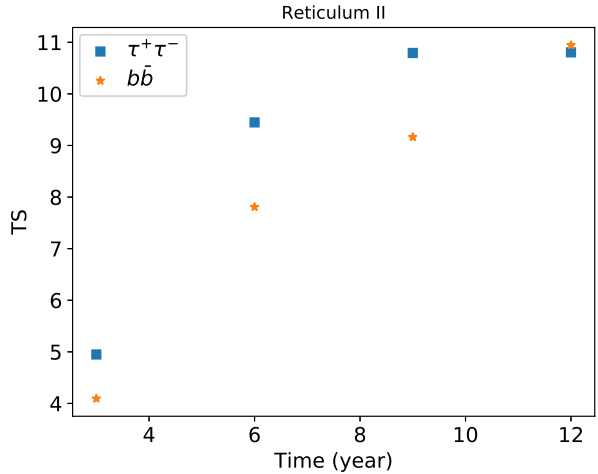}
\end{center}
\caption{The peak TS values of the possible $\gamma$-ray emission in the direction of Reticulum II for 3-, 6-, 9- and 12-year Fermi-LAT data. In the case of $\tau^{+}\tau^{-}$, the increasing behavior deviates significantly from being linear.}
\label{fig:r}
\end{figure}

\begin{figure*}[t]
\includegraphics[width=0.4\textwidth]{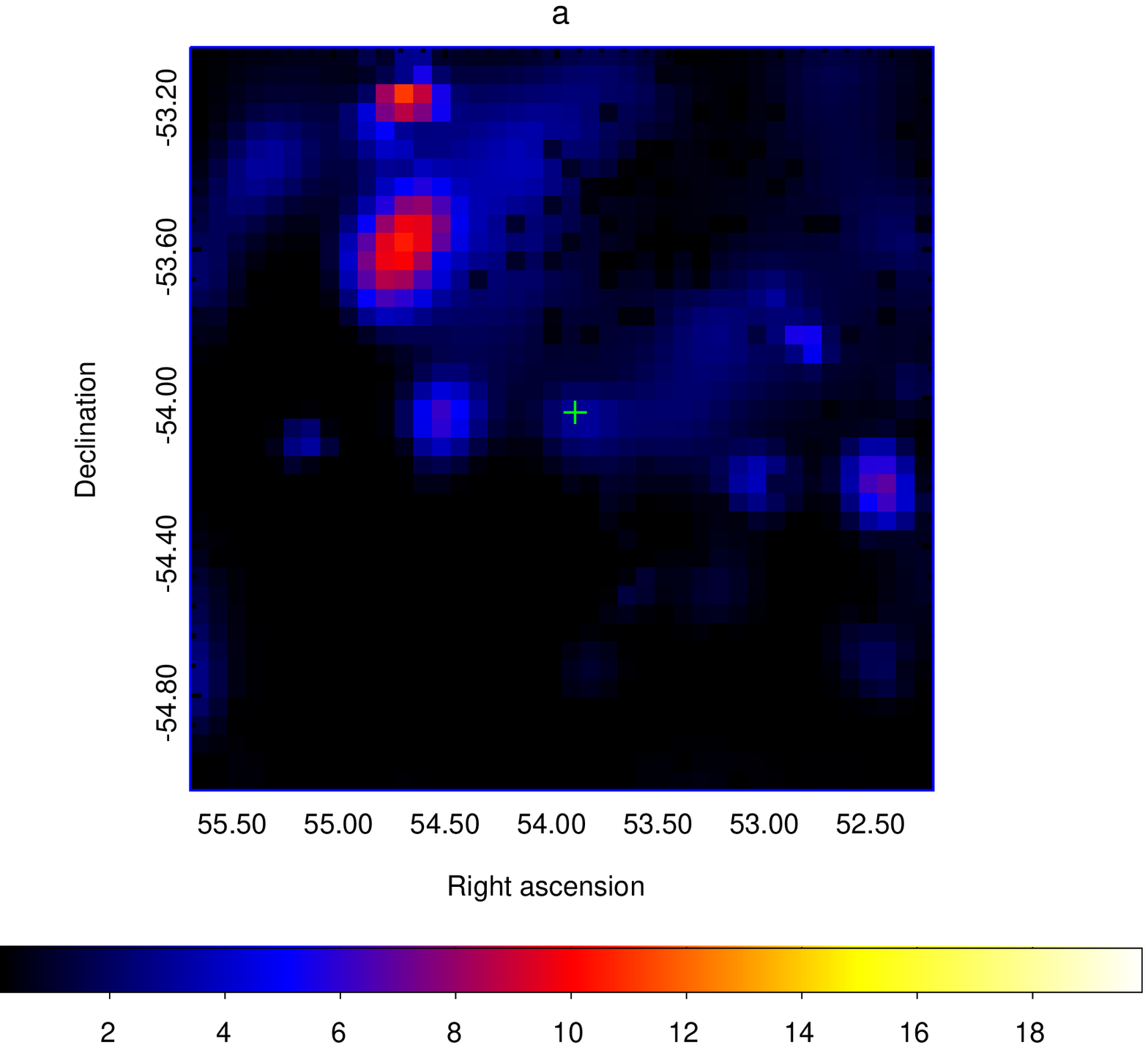}
\includegraphics[width=0.4\textwidth]{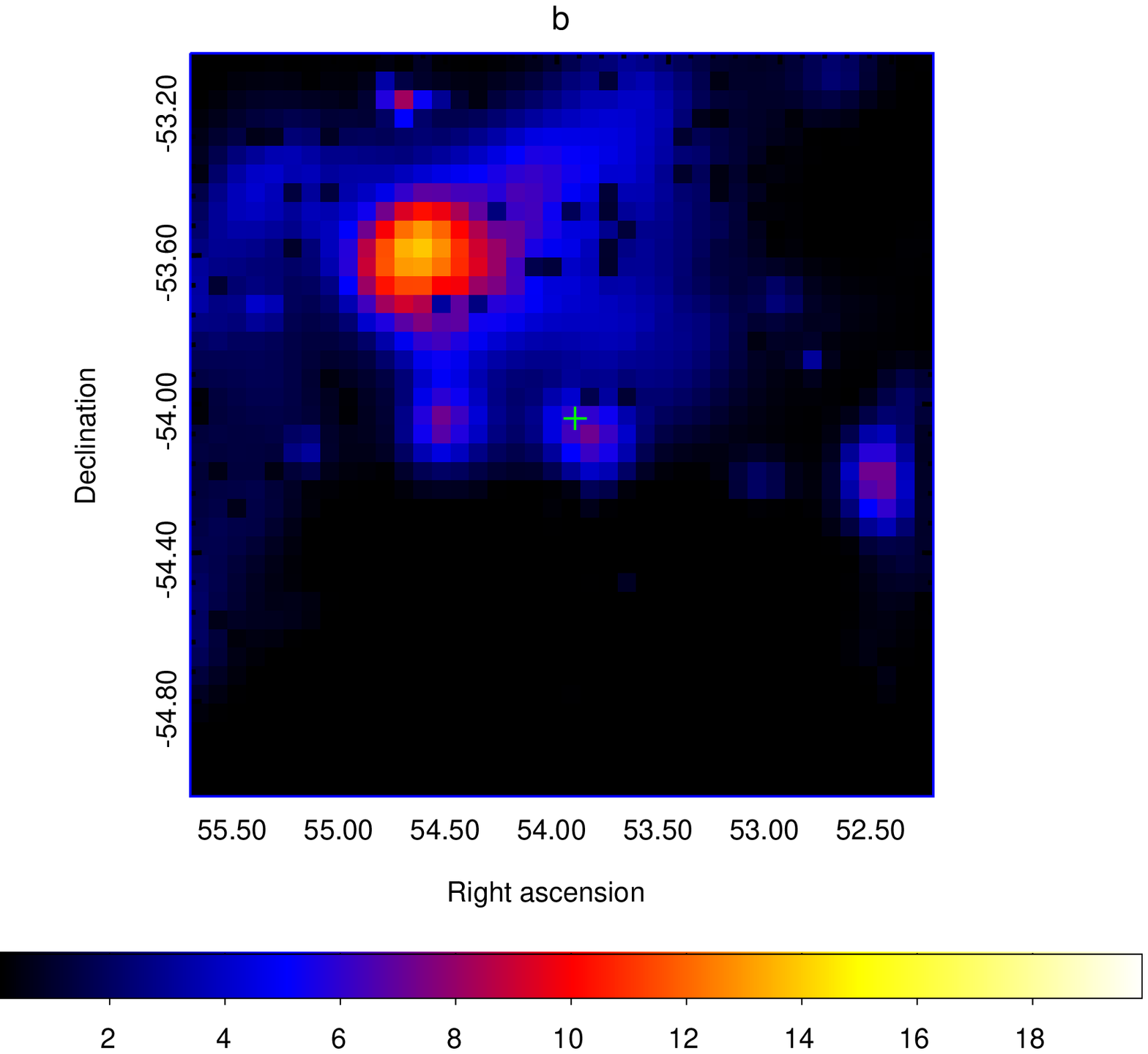}
\includegraphics[width=0.4\textwidth]{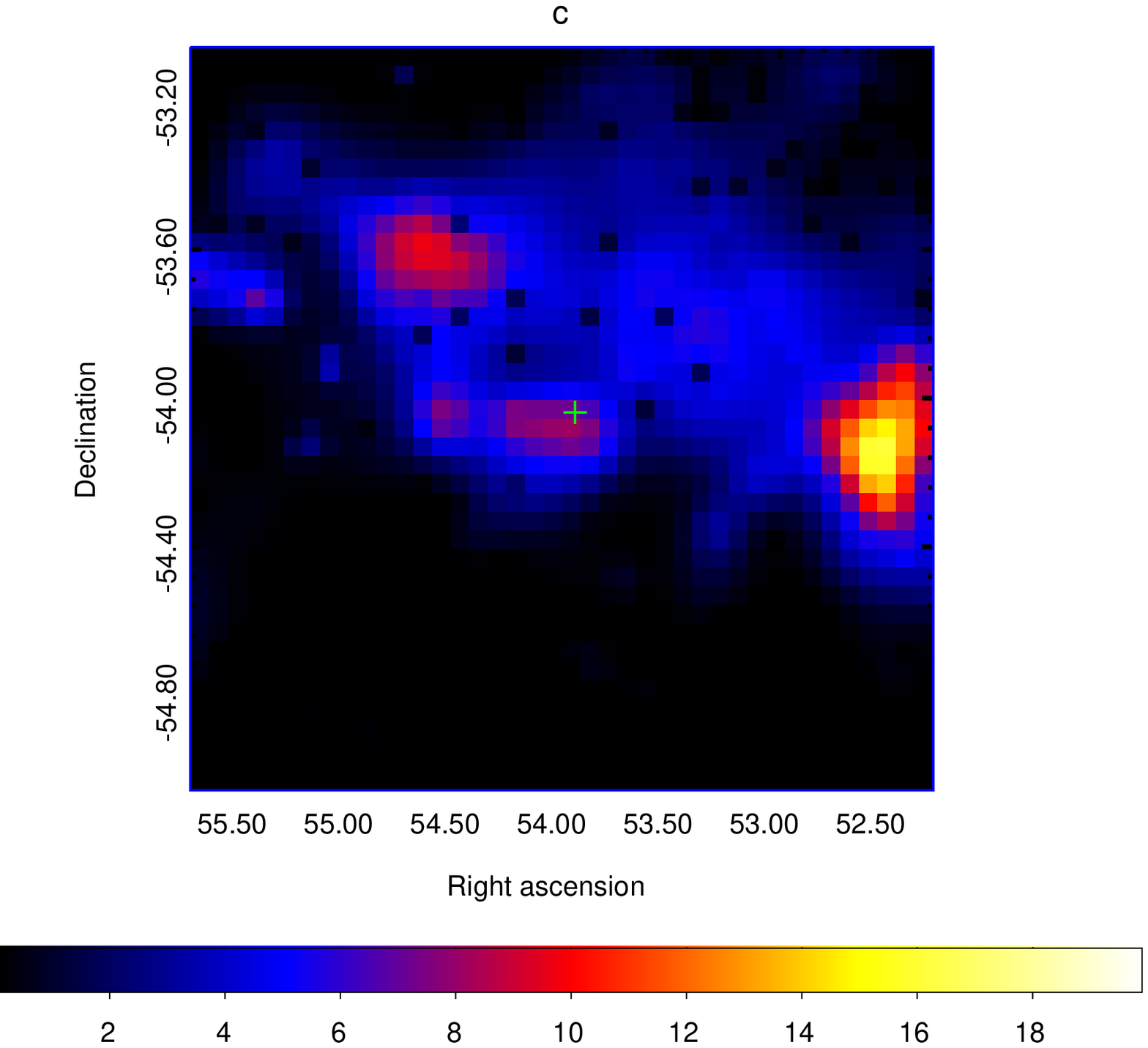}
\includegraphics[width=0.4\textwidth]{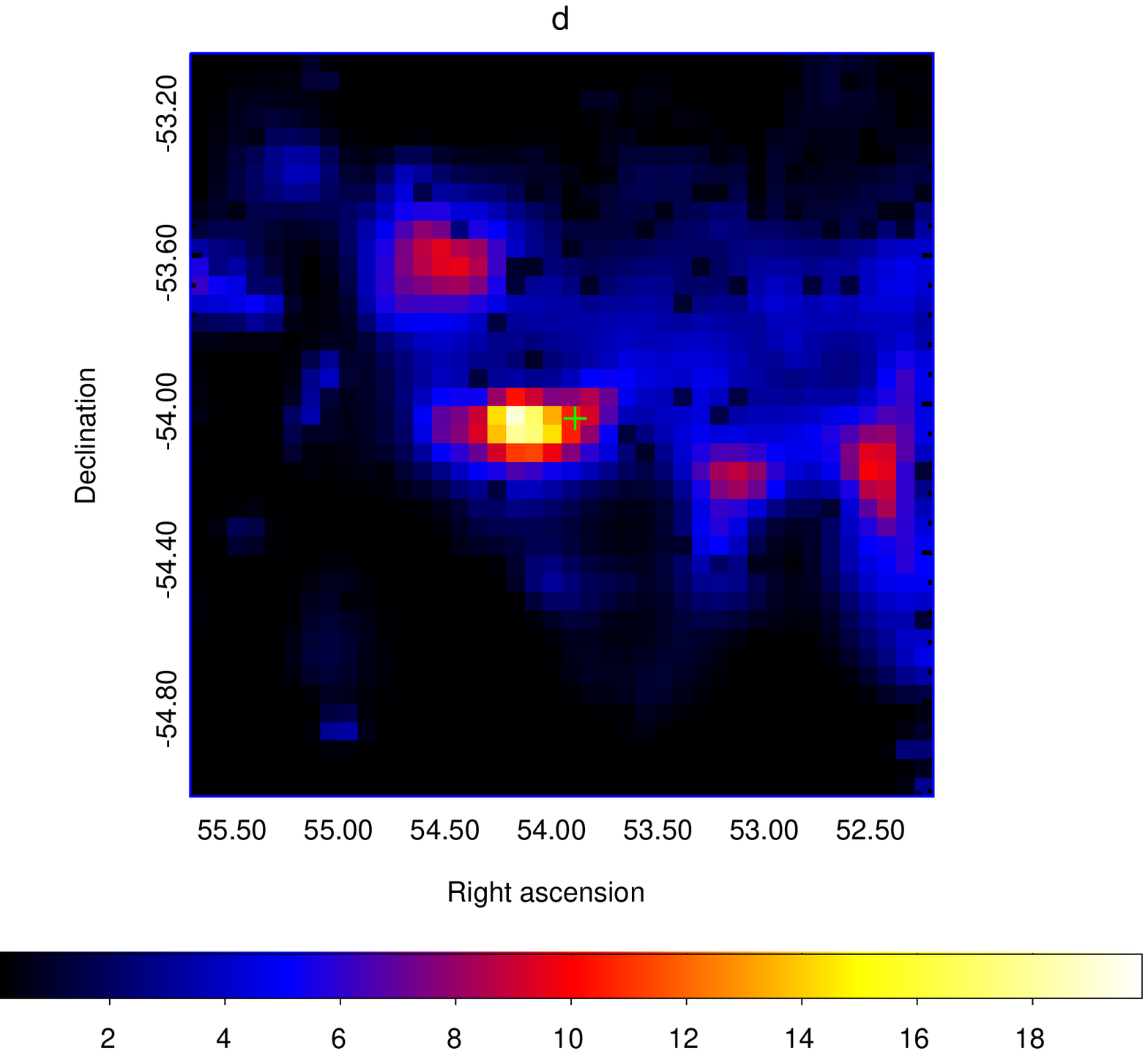}
\caption{$2^\circ\times2^\circ$ TS maps centered on Reticulum II with pixel size of $0.05^\circ$. The green cross symbol represents the optical position of Reticulum II. The a, b, c and d panels are for 3-, 6-, 9- and 12-year Fermi-LAT data, respectively.}	
\label{fig:rt}
\end{figure*}

\begin{figure}[t]
\begin{center}
\includegraphics[width=0.5\textwidth]{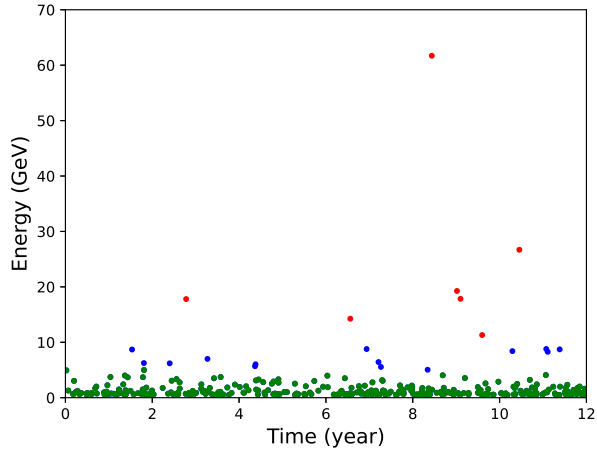}
\end{center}
\caption{The arrival times of the high energy $\gamma$-rays in the direction (i.e., within 0.5 degree) of Reticulum II. In the first eight years after the launch of Fermi-LAT, there were just two $\gamma$-rays with energies above 10 GeV were received. While in the recent 4 years, there were five such events. The filled circles in red (blue) are for photons with energies above $10$ GeV (above 5 GeV but below 10GeV).}
\label{fig:cts}
\end{figure}

\begin{figure}[!t]
\begin{center}
\includegraphics[width=0.5\textwidth]{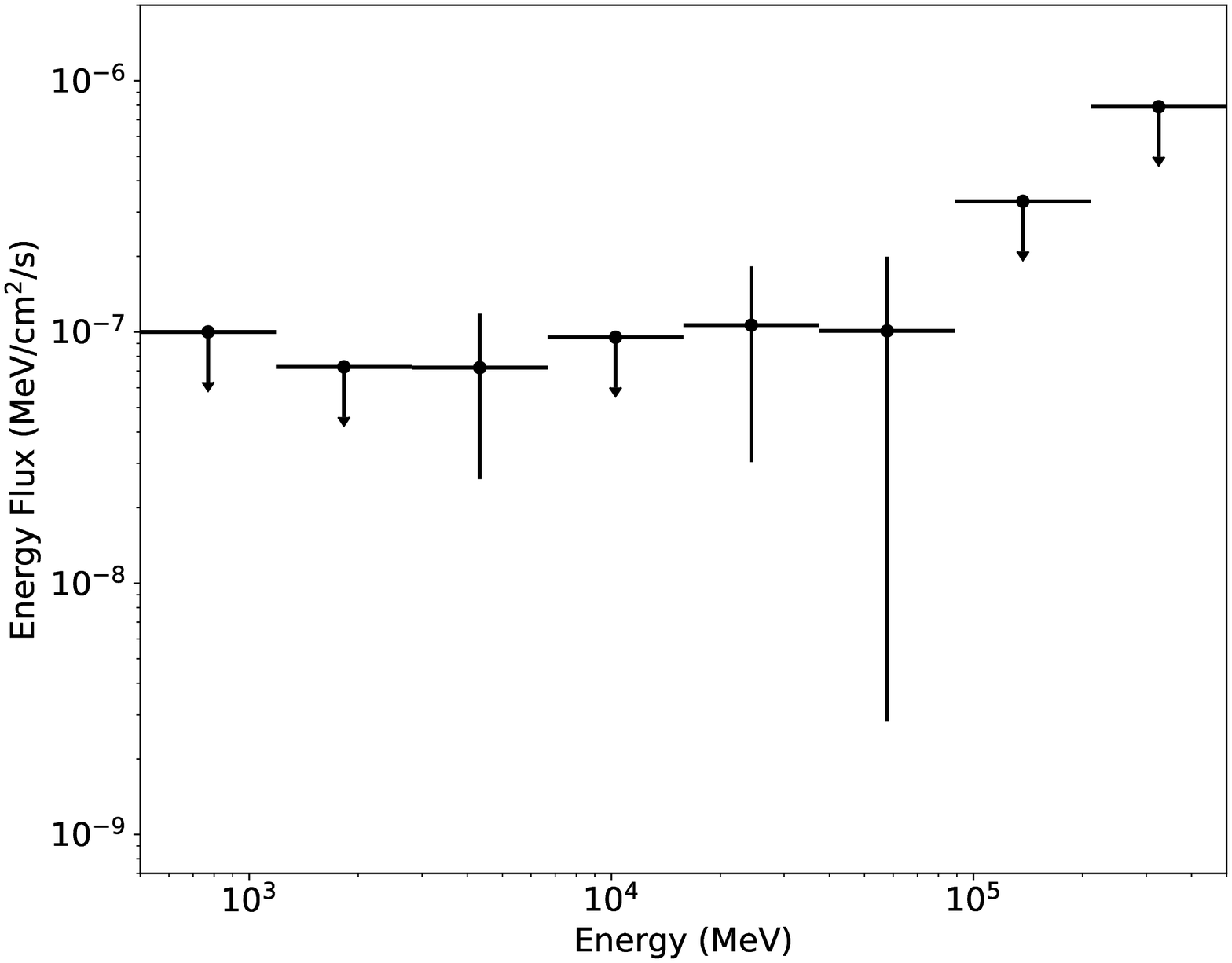}
\end{center}
\caption{Spectral energy distributions (SEDs) of the source A. Upper limits at a 95\% confidence level are derived when the TS values for the data points are lower than 4.}
\label{fig:sed}
\end{figure}

\section{Examining the three dSphs with weak $\gamma$-ray excesses}
\label{sec:ret2}

\subsection{Reticulum II}

The peak TS value of Reticulum II is $\sim10.8$ at $m_\chi \sim 36$ GeV for $\chi\chi\rightarrow\tau^{+}\tau^{-}$, while for $\chi\chi\rightarrow b\bar{b}$ the largest TS is $\sim10.9$ corresponding to a DM mass of $m_\chi \sim 300$ GeV. 
We notice that the TS value obtained here is lower than L18.
Performing the analysis of Reticulum II with data of different lengths, we find the TS value of the excess does not keep increasing with the data accumulation in the last three years for $\tau^+\tau^-$ channel (see Fig.~\ref{fig:r}). 
Furthermore, the best-fit DM masses are also inconsistent with the previous works \cite{gs15ret2, hooper15ret2,Drlica-Wagner:2015xua, fermi2016dsph,2018PhRvD..97l2001L}. They are generally larger than those in L18 (i.e., $\sim$90 GeV vs. $\sim$300 GeV for $b\bar{b}$, $\sim$16 GeV vs. $\sim$36 GeV for $\tau^+\tau^-$). This result challenges the DM model, since the spectrum of a DM signal should not change with time and thus the obtained DM mass will keep invariant in different epochs. 
However, since the signal is weak, the large uncertainty of the parameter may also account for this inconsistency.
We further examine why the fitting based on the 12-year data gives a higher DM mass by looking at individual photons from the direction of Reticulum II. Fig.~\ref{fig:cts} shows the energies and arrival times of the photons within the 0.5$^\circ$ region of Reticulum II. We can see that in recent four years, high energy ($>$10 GeV) photons surrounding Reticulum II increase remarkably, indicating a possible outburst of high energy photons. Such an enhancement is unexpected for the dark matter model but natural for the astrophysical process.

A TS map can provide us an intuitive impression of the background residual and can tell us whether the excess is due to bad modeling of the background (e.g., contamination from surrounding point sources). We create a $2^{\circ} \times 2^{\circ}$ TS map (the energy range from 500 MeV to 500 GeV) surrounding the target with the tool {\tt gttsmap} for the 12 years of Fermi-LAT data. In this process, the model parameters of all the sources within ROI are fixed to their best-fit values obtained in the global fit. 
In the TS map of 12-year data (see bottom right panel of Fig.~\ref{fig:rt}), though there is a possible $\gamma$-ray excess around Reticulum II, the excess is actually offset from the position of the target source.
The localization analysis gives the best-fit coordinates of RA=${54.18}^{\circ}$ and Dec=${-54.06}^{\circ}$ with a 2$\sigma$ error radius of 0.05$^{\circ}$. With the optimized $\gamma$-ray position, the TS value becomes $\sim$ 22.0 and the spectral energy distribution (SED) is showed in Fig.~\ref{fig:sed}. However, the Reticulum II is 0.15$^{\circ}$ offset from the best-fit position of the excess, significantly outside the error circle. If a new point source (denoted as source A) is added into background at the best-fit position, the TS of Reticulum II reduces to $\sim2.5$. 
The 3 -, 6 -, and 9-year TS maps are also shown in Fig.~\ref{fig:rt}. In general, there are weak excesses near the target source. But the positions do not well coincide with Reticulum II.

We also model the dSph with a spatially extended Navarro-Frenk-White (NFW) DM density profile \cite{Navarro:1996gj} and set the scale radius $r_s$ to 1 kpc. The $r_s$ corresponds to a angular extension of $<0.1^\circ$, which also does not cover the excess position A. The peak TS value with the extended spatial model is $\sim$ 10.0 ~(11.4) at ${m_{\chi}} \sim$ 14~(175) GeV for $\chi\chi\rightarrow\tau^{+}\tau^{-}~ (b\bar{b})$. 
Considering the source A is offset from Reticulum II, if not a background fluctuation, it may relate to an unidentified astrophysical $\gamma$-ray source. We attempt to identify the counterparts of this tentative source in other wavelengths but fail to find a suitable candidate.

We note that for this source the PL model gives a similar TS with the DM models (see Table \ref{tab:12dsph}). Together with the above-mentioned facts (i.e., the decline of the TS value, the sudden increase of high energy photons and the excess is positionally offset from Reticulum II), we conclude that the previously announced \cite{gs15ret2, hooper15ret2,Drlica-Wagner:2015xua, fermi2016dsph,2018PhRvD..97l2001L} weak $\gamma$-ray excess towards the direction of Reticulum II should have an astrophysical origin, irrelevant to the dark matter inside this dSph. This finding cautions us that a very careful background subtraction is important for searches of the  dark matter-induced faint $\gamma$-ray emissions from dSphs.

\subsection{Bootes II \& Willman 1} 

\begin{figure*}[t]
\includegraphics[width=0.45\textwidth]{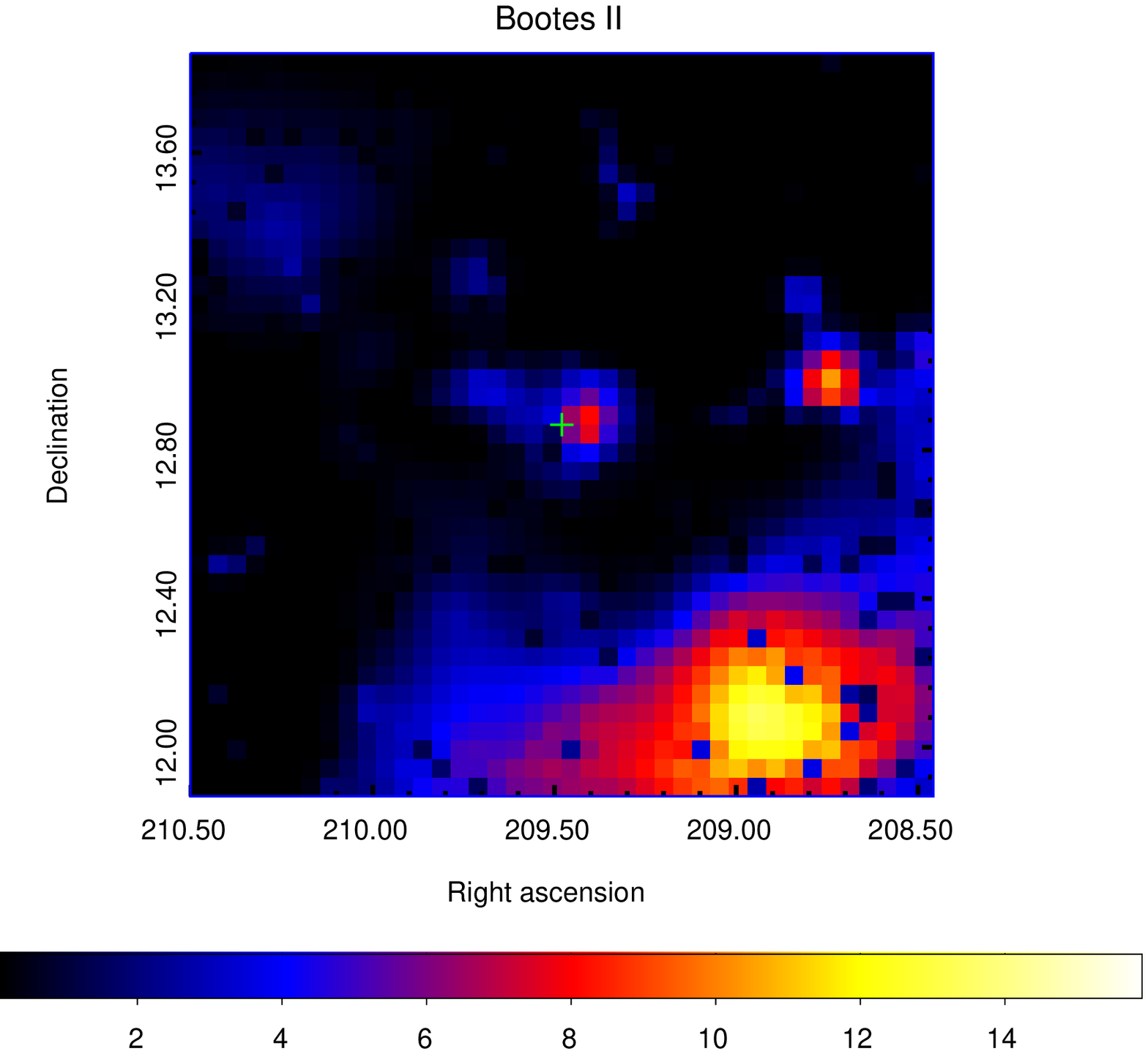}
\includegraphics[width=0.45\textwidth]{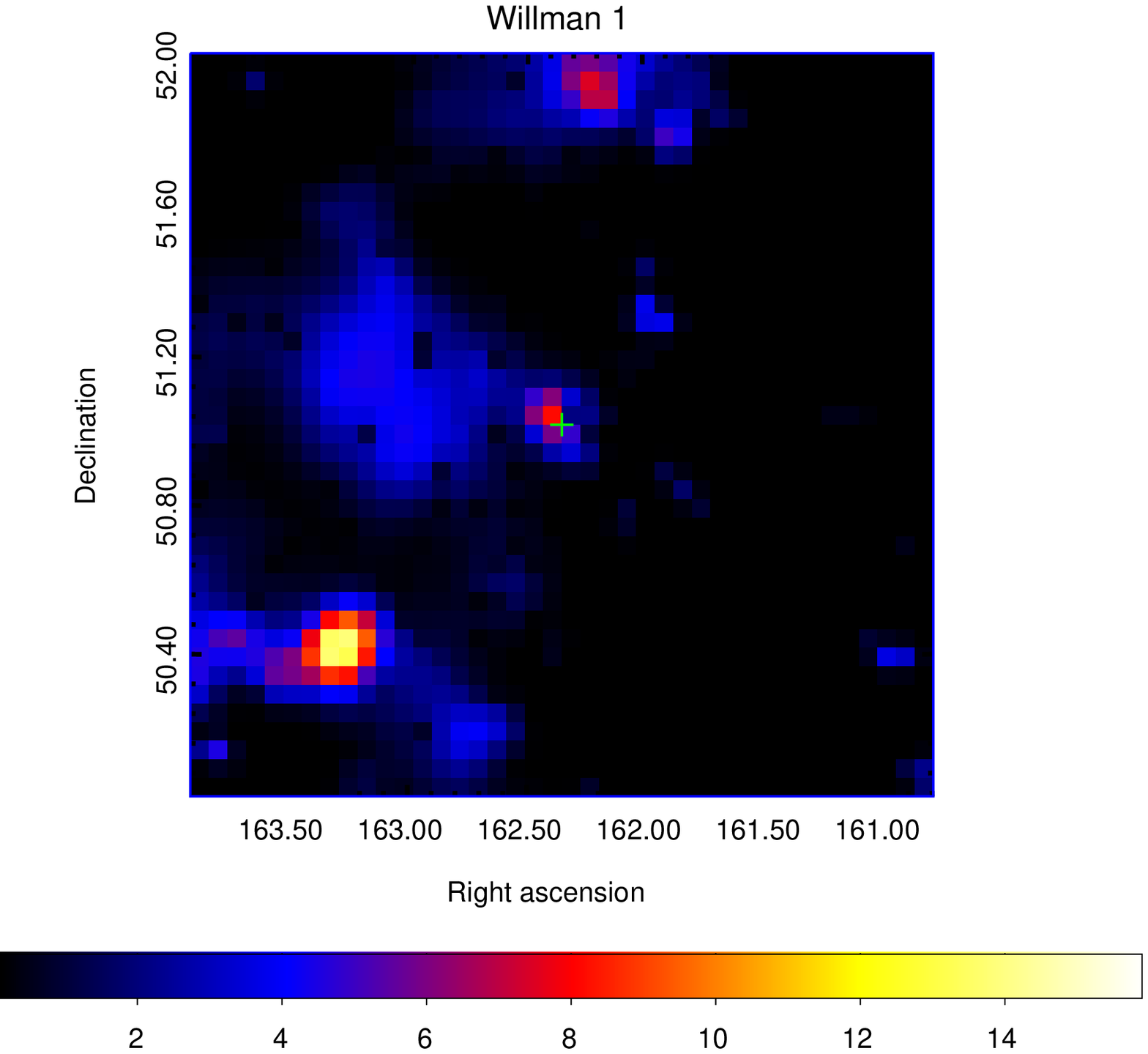}
\caption{The left panel is the residual TS maps of $2^{\circ} \times 2^{\circ}$ with $0.05^{\circ}$ per pixel centered at Bootes II and the optical position are marked in green cross symbol.
The right panel is the residual TS maps of $2^{\circ} \times 2^{\circ}$ with $0.05^{\circ}$ per pixel centered at Willman 1. The optical position of Willman 1 is marked with a green cross symbol.}
\label{fig:bs}
\end{figure*}

Other sources having mild TS values in our analyses are Bootes II and Willman 1. These two dSphs are both for the first time being reported to have weak $\gamma$-ray excesses. We generate residual TS maps of them, which are shown in Fig. \ref{fig:bs}. Very weak excesses actually appear around the two sources, however no more conclusive information can be drawn from the TS maps.

If modeled as DM signals, for Bootes II, the largest TS value is about 6.4 (5.4) for $\chi\chi\rightarrow\tau^{+}\tau^{-}$ ($b\bar{b}$) at $m_\chi \sim 14$ GeV (100 GeV). To reproduce the signal, a ${\left<{\sigma}v\right>}_{\chi\chi\rightarrow\tau^{+}\tau^{-}} \sim$ 4.1 $\times {10}^{-26}$ $\rm cm^3$$\rm s^{-1}$ (${\left<{\sigma}v\right>}_{\chi\chi\rightarrow b\bar{b}} \sim$ 1.9 $\times {10}^{-25}$ $\rm cm^3$$\rm s^{-1}$) is needed adopting the empirical J-factor. For Willman 1, the peak TS value is $\sim$ 7.3 at ${m_{\chi}} \sim$ 80 GeV for $\chi\chi\rightarrow\tau^{+}\tau^{-}$, whereas for $\chi\chi\rightarrow b\bar{b}$ it is $\sim 7.6$ at ${m_{\chi}} \sim$ 500GeV. The requried cross section is ${\left<{\sigma}v\right>}_{\chi\chi\rightarrow\tau^{+}\tau^{-}} \sim$ 2.6 $\times {10}^{-25}$ $\rm cm^3$$\rm s^{-1}$ or ${\left<{\sigma}v\right>}_{\chi\chi\rightarrow b\bar{b}} \sim$ 1.0 $\times {10}^{-24}$ $\rm cm^3$$\rm s^{-1}$. 
We notice that for both sources the derived cross sections $\left<{\sigma}v\right>$ are all excluded by \cite{fermi2016dsph}, where the upper limits are derived based on kinematically determined J-factors. In addition, several dSphs in our sample have J-factors comparable to or even larger than Bootes II and Willman 1, but do not show any excesses over the background. The results suggest that the weak excesses may be not from the DM annihilation, unless the real J-factors of the two dSphs are significantly larger than our simple estimates. Please note that the large uncertainty of the derived $\left<{\sigma}v\right>$ has not been taken into account in the comparison. Moreover, it should be noted that Willman 1 may be a tidal disruption system according the kinematic and photometric observations \cite{Willman_2011}. We further model the targets with spatially extended NFW DM density profiles (rather than point-like sources). But the obtained TS values are not larger than the above point source analysis.

Though it is hard to claim a $\gamma$-ray signal at this stage due to the small TS values, for completeness we still search for the possible astrophysical counterparts of these two excesses at other wavelengths. We find no 4FGL-DR2 sources are located within 0.5$^{\circ}$ of the targets. In the BZCAT \cite{2009A&A...495..691M}, CRATES \cite{2007ApJS..171...61H}, CGraBS \cite{2008ApJS..175...97H} and WISE blazar candidate catalogs \cite{2014ApJS..215...14D}, no potential $\gamma$-ray emitters are found to be within 0.5$^{\circ}$ of these two dSphs.

\section{Summary and Discussion} 

The Milky Way dwarf spheroidal galaxies are dominated by DM and are ideal objects for indirect detection of DM signals. 
Currently, no significant $\gamma$-ray signals have been found in dSphs and people only reported weak possible 'excesses' from some dSphs \cite{gs15ret2, hooper15ret2,Drlica-Wagner:2015xua, fermi2016dsph, 2018PhRvD..97l2001L, li16dsph, Bhattacharjee:2018xem}. 
No matter due to DM annihilation or astrophysical process, the discovery of $\gamma$-ray signals in dSphs is an important progress in our understanding of this type of objects, which motivates ongoing efforts to search for such emission.
In this paper, we performed a comprehensive $\gamma$-ray analysis on the Fermi-LAT observation to the 12 nearest dSphs (including candidates), which is an update of our previous work \cite{2018PhRvD..97l2001L}. We use longer dataset and the latest Fermi-LAT software and the latest background models to carry out the analysis aiming to examine the tentative $\gamma$-ray excesses reported previously.

We find no significant ($>5\sigma$) $\gamma$-ray signals but just very weak $\gamma$-ray excesses in the direction of Reticulum II, Bootes II and Willman 1, for which the largest TS values are $\sim$10.9, $\sim$ 6.4 and $\sim$7.6 in the scenario of DM annihilation, respectively. In the past, Reticulum II had attracted wide attention since  it displays the most significant, though still weak, $\gamma$-ray signal among all dSphs/candidates.  
However, with the 12-year Fermi-LAT data we find that the position of the weak excess is significantly away from the center of Reticulum II any more, which strongly suggests that the excess is irrelevant to the dark matter annihilation. Other clues disfavoring a DM origin of the excess include: the decline of the TS value in the new analysis (compared to previous works), a sudden increase of the photons above 10 GeV in recent years. Reticulum II has a half-light radius of $r_{\rm h}\approx 55$ pc. The offset by an angular of $0.15^\circ$ corresponds to a radius of $\sim 83$ pc, which is comparable with $r_{\rm h}$. We suggest that the GeV emission with a TS value of $\sim 22.0$ has the regular astrophysical origin. It is however unclear whether this ``new" source is within Reticulum II or not. 

For Bootes II and Willman 1, they are the first time being reported to have mild $\gamma$-ray excesses. Due to the faintness of the excesses, we can not determine their origin at this time. They may be from astrophysical processes, DM annihilations or, most probably, background fluctuations. Even so, considering the dark matter origin of the GeV excess from Reticulum II is not supported by observation, at present these two targets are the most promising ones. 
It is therefore worth to pay more attention to them. The Fermi-LAT and other operating/future $\gamma$-ray telescopes \cite{gamma400,zhang14herd}{}\footnote{\url{http://www2.yukawa.kyoto-u.ac.jp/~mmgw2019/slide/5th/Fan.pdf}} may reveal the nature of these tentative signals in the future.

\begin{acknowledgments}
 This work is supported by the National Natural Science Foundation of China (No. U1738210 and No.11921003).
\end{acknowledgments}

\begin{thebibliography}{60}%
\makeatletter
\providecommand \@ifxundefined [1]{%
 \@ifx{#1\undefined}
}%
\providecommand \@ifnum [1]{%
 \ifnum #1\expandafter \@firstoftwo
 \else \expandafter \@secondoftwo
 \fi
}%
\providecommand \@ifx [1]{%
 \ifx #1\expandafter \@firstoftwo
 \else \expandafter \@secondoftwo
 \fi
}%
\providecommand \natexlab [1]{#1}%
\providecommand \enquote  [1]{``#1''}%
\providecommand \bibnamefont  [1]{#1}%
\providecommand \bibfnamefont [1]{#1}%
\providecommand \citenamefont [1]{#1}%
\providecommand \href@noop [0]{\@secondoftwo}%
\providecommand \href [0]{\begingroup \@sanitize@url \@href}%
\providecommand \@href[1]{\@@startlink{#1}\@@href}%
\providecommand \@@href[1]{\endgroup#1\@@endlink}%
\providecommand \@sanitize@url [0]{\catcode `\\12\catcode `\$12\catcode
  `\&12\catcode `\#12\catcode `\^12\catcode `\_12\catcode `\%12\relax}%
\providecommand \@@startlink[1]{}%
\providecommand \@@endlink[0]{}%
\providecommand \url  [0]{\begingroup\@sanitize@url \@url }%
\providecommand \@url [1]{\endgroup\@href {#1}{\urlprefix }}%
\providecommand \urlprefix  [0]{URL }%
\providecommand \Eprint [0]{\href }%
\providecommand \doibase [0]{http://dx.doi.org/}%
\providecommand \selectlanguage [0]{\@gobble}%
\providecommand \bibinfo  [0]{\@secondoftwo}%
\providecommand \bibfield  [0]{\@secondoftwo}%
\providecommand \translation [1]{[#1]}%
\providecommand \BibitemOpen [0]{}%
\providecommand \bibitemStop [0]{}%
\providecommand \bibitemNoStop [0]{.\EOS\space}%
\providecommand \EOS [0]{\spacefactor3000\relax}%
\providecommand \BibitemShut  [1]{\csname bibitem#1\endcsname}%
\let\auto@bib@innerbib\@empty
\bibitem [{\citenamefont {Ade}\ \emph {{\it et~al.}}(2016)\citenamefont {Ade}
  {\it et~al.}}]{Ade:2015xua}%
  \BibitemOpen
  \bibfield  {author} {\bibinfo {author} {\bibfnamefont {P.~A.~R.}\
  \bibnamefont {Ade}}  {\it et~al.} (\bibinfo {collaboration} {Planck}
  Collaboration),\ }\bibfield  {title} {\enquote {\bibinfo {title} {{Planck
  2015 results. XIII. Cosmological parameters}},}\ }\href {\doibase
  10.1051/0004-6361/201525830} {\bibfield  {journal} {\bibinfo  {journal}
  {Astron. Astrophys.}\ }\textbf {\bibinfo {volume} {594}},\ \bibinfo {pages}
  {A13} (\bibinfo {year} {2016})},\ \Eprint
  {http://arxiv.org/abs/1502.01589}{arXiv:1502.01589}\BibitemShut {NoStop}%
\bibitem [{\citenamefont {Jungman}\ \emph {{\it et~al.}}(1996)\citenamefont
  {Jungman}, \citenamefont {Kamionkowski},\ and\ \citenamefont
  {Griest}}]{Jungman:1995df}%
  \BibitemOpen
  \bibfield  {author} {\bibinfo {author} {\bibfnamefont {G.}~\bibnamefont
  {Jungman}}, \bibinfo {author} {\bibfnamefont {M.}~\bibnamefont
  {Kamionkowski}}, and\ \bibinfo {author} {\bibfnamefont {K.}~\bibnamefont
  {Griest}},\ }\bibfield  {title} {\enquote {\bibinfo {title} {{Supersymmetric
  dark matter}},}\ }\href {\doibase 10.1016/0370-1573(95)00058-5} {\bibfield
  {journal} {\bibinfo  {journal} {Phys. Rept.}\ }\textbf {\bibinfo {volume}
  {267}},\ \bibinfo {pages} {195} (\bibinfo {year} {1996})},\ \Eprint
  {http://arxiv.org/abs/hep-ph/9506380}{arXiv:hep-ph/9506380}\BibitemShut
  {NoStop}%
\bibitem [{\citenamefont {Bertone}\ \emph {{\it et~al.}}(2005)\citenamefont
  {Bertone}, \citenamefont {Hooper},\ and\ \citenamefont
  {Silk}}]{Bertone:2004pz}%
  \BibitemOpen
  \bibfield  {author} {\bibinfo {author} {\bibfnamefont {G.}~\bibnamefont
  {Bertone}}, \bibinfo {author} {\bibfnamefont {D.}~\bibnamefont {Hooper}},
  and\ \bibinfo {author} {\bibfnamefont {J.}~\bibnamefont {Silk}},\ }\bibfield
  {title} {\enquote {\bibinfo {title} {{Particle dark matter: Evidence,
  candidates and constraints}},}\ }\href {\doibase
  10.1016/j.physrep.2004.08.031} {\bibfield  {journal} {\bibinfo  {journal}
  {Phys. Rept.}\ }\textbf {\bibinfo {volume} {405}},\ \bibinfo {pages} {279}
  (\bibinfo {year} {2005})},\ \Eprint
  {http://arxiv.org/abs/hep-ph/0404175}{arXiv:hep-ph/0404175}\BibitemShut
  {NoStop}%
\bibitem [{\citenamefont {Hooper}\ and\ \citenamefont
  {Profumo}(2007)}]{Hooper:2007qk}%
  \BibitemOpen
  \bibfield  {author} {\bibinfo {author} {\bibfnamefont {D.}~\bibnamefont
  {Hooper}} and\ \bibinfo {author} {\bibfnamefont {S.}~\bibnamefont
  {Profumo}},\ }\bibfield  {title} {\enquote {\bibinfo {title} {{Dark matter
  and collider phenomenology of universal extra dimensions}},}\ }\href
  {\doibase 10.1016/j.physrep.2007.09.003} {\bibfield  {journal} {\bibinfo
  {journal} {Phys. Rept.}\ }\textbf {\bibinfo {volume} {453}},\ \bibinfo
  {pages} {29} (\bibinfo {year} {2007})},\ \Eprint
  {http://arxiv.org/abs/hep-ph/0701197}{arXiv:hep-ph/0701197}\BibitemShut
  {NoStop}%
\bibitem [{\citenamefont {Feng}(2010)}]{Feng:2010gw}%
  \BibitemOpen
  \bibfield  {author} {\bibinfo {author} {\bibfnamefont {J.~L.}\ \bibnamefont
  {Feng}},\ }\bibfield  {title} {\enquote {\bibinfo {title} {{Dark Matter
  Candidates from Particle Physics and Methods of Detection}},}\ }\href
  {\doibase 10.1146/annurev-astro-082708-101659} {\bibfield  {journal}
  {\bibinfo  {journal} {Ann. Rev. Astron. Astrophys.}\ }\textbf {\bibinfo
  {volume} {48}},\ \bibinfo {pages} {495} (\bibinfo {year} {2010})},\ \Eprint
  {http://arxiv.org/abs/1003.0904}{arXiv:1003.0904}\BibitemShut {NoStop}%
\bibitem [{\citenamefont {Evans}\ \emph {{\it et~al.}}(2004)\citenamefont
  {Evans}, \citenamefont {Ferrer},\ and\ \citenamefont
  {Sarkar}}]{PhysRevD.69.123501}%
  \BibitemOpen
  \bibfield  {author} {\bibinfo {author} {\bibfnamefont {N.~W.}\ \bibnamefont
  {Evans}}, \bibinfo {author} {\bibfnamefont {F.}~\bibnamefont {Ferrer}}, and\
  \bibinfo {author} {\bibfnamefont {S.}~\bibnamefont {Sarkar}},\ }\bibfield
  {title} {\enquote {\bibinfo {title} {A travel guide to the dark matter
  annihilation signal},}\ }\href {\doibase 10.1103/PhysRevD.69.123501}
  {\bibfield  {journal} {\bibinfo  {journal} {Phys. Rev. D}\ }\textbf {\bibinfo
  {volume} {69}},\ \bibinfo {pages} {123501} (\bibinfo {year}
  {2004})}\BibitemShut {NoStop}%
\bibitem [{\citenamefont {Atwood}\ \emph {{\it et~al.}}(2009)\citenamefont
  {Atwood} {\it et~al.}}]{atwood09lat}%
  \BibitemOpen
  \bibfield  {author} {\bibinfo {author} {\bibfnamefont {W.~B.}\ \bibnamefont
  {Atwood}}  {\it et~al.} (\bibinfo {collaboration} {Fermi-LAT}
  Collaboration),\ }\bibfield  {title} {\enquote {\bibinfo {title} {{The Large
  Area Telescope on the Fermi Gamma-ray Space Telescope Mission}},}\ }\href
  {\doibase 10.1088/0004-637X/697/2/1071} {\bibfield  {journal} {\bibinfo
  {journal} {Astrophys. J.}\ }\textbf {\bibinfo {volume} {697}},\ \bibinfo
  {pages} {1071} (\bibinfo {year} {2009})},\ \Eprint
  {http://arxiv.org/abs/0902.1089}{arXiv:0902.1089}\BibitemShut {NoStop}%
\bibitem [{\citenamefont {Chang}\ \emph {{\it et~al.}}(2017)\citenamefont
  {Chang} {\it et~al.}}]{DAMPE:2017}%
  \BibitemOpen
  \bibfield  {author} {\bibinfo {author} {\bibfnamefont {J.}~\bibnamefont
  {Chang}}  {\it et~al.} (\bibinfo {collaboration} {DAMPE} Collaboration),\
  }\bibfield  {title} {\enquote {\bibinfo {title} {{The DArk Matter Particle
  Explorer mission}},}\ }\href {\doibase 10.1016/j.astropartphys.2017.08.005}
  {\bibfield  {journal} {\bibinfo  {journal} {Astropart. Phys.}\ }\textbf
  {\bibinfo {volume} {95}},\ \bibinfo {pages} {6} (\bibinfo {year} {2017})},\
  \Eprint {http://arxiv.org/abs/1706.08453}{arXiv:1706.08453}\BibitemShut
  {NoStop}%
\bibitem [{\citenamefont {Ambrosi}\ \emph {{\it et~al.}}(2017)\citenamefont
  {Ambrosi} {\it et~al.}}]{Chang:2017n}%
  \BibitemOpen
  \bibfield  {author} {\bibinfo {author} {\bibfnamefont {G.}~\bibnamefont
  {Ambrosi}}  {\it et~al.} (\bibinfo {collaboration} {DAMPE} Collaboration),\
  }\bibfield  {title} {\enquote {\bibinfo {title} {{Direct detection of a break
  in the teraelectronvolt cosmic-ray spectrum of electrons and positrons}},}\
  }\href {\doibase 10.1038/nature24475} {\bibfield  {journal} {\bibinfo
  {journal} {Nature}\ }\textbf {\bibinfo {volume} {552}},\ \bibinfo {pages}
  {63} (\bibinfo {year} {2017})},\ \Eprint
  {http://arxiv.org/abs/1711.10981}{arXiv:1711.10981}\BibitemShut {NoStop}%
\bibitem [{\citenamefont {Lake}(1990)}]{Lake:1990du}%
  \BibitemOpen
  \bibfield  {author} {\bibinfo {author} {\bibfnamefont {G.}~\bibnamefont
  {Lake}},\ }\bibfield  {title} {\enquote {\bibinfo {title} {{Detectability of
  gamma-rays from clumps of dark matter}},}\ }\href {\doibase 10.1038/346039a0}
  {\bibfield  {journal} {\bibinfo  {journal} {Nature}\ }\textbf {\bibinfo
  {volume} {346}},\ \bibinfo {pages} {39} (\bibinfo {year} {1990})}\BibitemShut
  {NoStop}%
\bibitem [{\citenamefont {Baltz}\ and\ \citenamefont
  {Wai}(2004)}]{Baltz:2004bb}%
  \BibitemOpen
  \bibfield  {author} {\bibinfo {author} {\bibfnamefont {E.~A.}\ \bibnamefont
  {Baltz}} and\ \bibinfo {author} {\bibfnamefont {L.}~\bibnamefont {Wai}},\
  }\bibfield  {title} {\enquote {\bibinfo {title} {{Diffuse inverse Compton and
  synchrotron emission from dark matter annihilations in galactic
  satellites}},}\ }\href {\doibase 10.1103/PhysRevD.70.023512} {\bibfield
  {journal} {\bibinfo  {journal} {Phys. Rev. D}\ }\textbf {\bibinfo {volume}
  {70}},\ \bibinfo {pages} {023512} (\bibinfo {year} {2004})},\ \Eprint
  {http://arxiv.org/abs/astro-ph/0403528}{arXiv:astro-ph/0403528}\BibitemShut
  {NoStop}%
\bibitem [{\citenamefont {Strigari}(2013)}]{Strigari:2013iaa}%
  \BibitemOpen
  \bibfield  {author} {\bibinfo {author} {\bibfnamefont {L.~E.}\ \bibnamefont
  {Strigari}},\ }\bibfield  {title} {\enquote {\bibinfo {title} {{Galactic
  Searches for Dark Matter}},}\ }\href {\doibase 10.1016/j.physrep.2013.05.004}
  {\bibfield  {journal} {\bibinfo  {journal} {Phys. Rept.}\ }\textbf {\bibinfo
  {volume} {531}},\ \bibinfo {pages} {1} (\bibinfo {year} {2013})},\ \Eprint
  {http://arxiv.org/abs/1211.7090}{arXiv:1211.7090}\BibitemShut {NoStop}%
\bibitem [{\citenamefont {{Ackermann}}\ \emph {{\it et~al.}}(2011)\citenamefont
  {{Ackermann}}, \citenamefont {{Ajello}}, \citenamefont {{Albert}},
  \citenamefont {{Atwood}}, \citenamefont {{Baldini}}, \citenamefont
  {{Ballet}}, \citenamefont {{Barbiellini}}, \citenamefont {{Bastieri}},
  \citenamefont {{Bechtol}}, \citenamefont {{Bellazzini}}, \citenamefont
  {{Berenji}}, \citenamefont {{Blandford}}, \citenamefont {{Bloom}},
  \citenamefont {{Bonamente}}, \citenamefont {{Borgland}}, \citenamefont
  {{Bregeon}}, \citenamefont {{Brigida}}, \citenamefont {{Bruel}},
  \citenamefont {{Buehler}}, \citenamefont {{Burnett}}, \citenamefont
  {{Buson}}, \citenamefont {{Caliandro}}, \citenamefont {{Cameron}},
  \citenamefont {{Ca{\~n}adas}}, \citenamefont {{Caraveo}}, \citenamefont
  {{Casandjian}}, \citenamefont {{Cecchi}}, \citenamefont {{Charles}},
  \citenamefont {{Chekhtman}}, \citenamefont {{Chiang}}, \citenamefont
  {{Ciprini}}, \citenamefont {{Claus}}, \citenamefont {{Cohen-Tanugi}},
  \citenamefont {{Conrad}}, \citenamefont {{Cutini}}, \citenamefont {{de
  Angelis}}, \citenamefont {{de Palma}}, \citenamefont {{Dermer}},
  \citenamefont {{Digel}}, \citenamefont {{Do Couto E Silva}}, \citenamefont
  {{Drell}}, \citenamefont {{Drlica-Wagner}}, \citenamefont {{Falletti}},
  \citenamefont {{Favuzzi}}, \citenamefont {{Fegan}}, \citenamefont
  {{Ferrara}}, \citenamefont {{Fukazawa}}, \citenamefont {{Funk}},
  \citenamefont {{Fusco}}, \citenamefont {{Gargano}}, \citenamefont
  {{Gasparrini}}, \citenamefont {{Gehrels}}, \citenamefont {{Germani}},
  \citenamefont {{Giglietto}}, \citenamefont {{Giordano}}, \citenamefont
  {{Giroletti}}, \citenamefont {{Glanzman}}, \citenamefont {{Godfrey}},
  \citenamefont {{Grenier}}, \citenamefont {{Guiriec}}, \citenamefont
  {{Gustafsson}}, \citenamefont {{Hadasch}}, \citenamefont {{Hayashida}},
  \citenamefont {{Hays}}, \citenamefont {{Hughes}}, \citenamefont {{Jeltema}},
  \citenamefont {{J{\'o}hannesson}}, \citenamefont {{Johnson}}, \citenamefont
  {{Johnson}}, \citenamefont {{Kamae}}, \citenamefont {{Katagiri}},
  \citenamefont {{Kataoka}}, \citenamefont {{Kn{\"o}dlseder}}, \citenamefont
  {{Kuss}}, \citenamefont {{Lande}}, \citenamefont {{Latronico}}, \citenamefont
  {{Lionetto}}, \citenamefont {{Llena Garde}}, \citenamefont {{Longo}},
  \citenamefont {{Loparco}}, \citenamefont {{Lott}}, \citenamefont
  {{Lovellette}}, \citenamefont {{Lubrano}}, \citenamefont {{Madejski}},
  \citenamefont {{Mazziotta}}, \citenamefont {{McEnery}}, \citenamefont
  {{Mehault}}, \citenamefont {{Michelson}}, \citenamefont {{Mitthumsiri}},
  \citenamefont {{Mizuno}}, \citenamefont {{Monte}}, \citenamefont {{Monzani}},
  \citenamefont {{Morselli}}, \citenamefont {{Moskalenko}}, \citenamefont
  {{Murgia}}, \citenamefont {{Naumann-Godo}}, \citenamefont {{Norris}},
  \citenamefont {{Nuss}}, \citenamefont {{Ohsugi}}, \citenamefont {{Okumura}},
  \citenamefont {{Omodei}}, \citenamefont {{Orlando}}, \citenamefont {{Ormes}},
  \citenamefont {{Ozaki}}, \citenamefont {{Paneque}}, \citenamefont {{Parent}},
  \citenamefont {{Pesce-Rollins}}, \citenamefont {{Pierbattista}},
  \citenamefont {{Piron}}, \citenamefont {{Pivato}}, \citenamefont {{Porter}},
  \citenamefont {{Profumo}}, \citenamefont {{Rain{\`o}}}, \citenamefont
  {{Razzano}}, \citenamefont {{Reimer}}, \citenamefont {{Reimer}},
  \citenamefont {{Ritz}}, \citenamefont {{Roth}}, \citenamefont
  {{Sadrozinski}}, \citenamefont {{Sbarra}}, \citenamefont {{Scargle}},
  \citenamefont {{Schalk}}, \citenamefont {{Sgr{\`o}}}, \citenamefont
  {{Siskind}}, \citenamefont {{Spandre}}, \citenamefont {{Spinelli}},
  \citenamefont {{Strigari}}, \citenamefont {{Suson}}, \citenamefont
  {{Tajima}}, \citenamefont {{Takahashi}}, \citenamefont {{Tanaka}},
  \citenamefont {{Thayer}}, \citenamefont {{Thayer}}, \citenamefont
  {{Thompson}}, \citenamefont {{Tibaldo}}, \citenamefont {{Tinivella}},
  \citenamefont {{Torres}}, \citenamefont {{Troja}}, \citenamefont
  {{Uchiyama}}, \citenamefont {{Vandenbroucke}}, \citenamefont {{Vasileiou}},
  \citenamefont {{Vianello}}, \citenamefont {{Vitale}}, \citenamefont
  {{Waite}}, \citenamefont {{Wang}}, \citenamefont {{Winer}}, \citenamefont
  {{Wood}}, \citenamefont {{Wood}}, \citenamefont {{Yang}}, \citenamefont
  {{Zimmer}}, \citenamefont {{Kaplinghat}},\ and\ \citenamefont
  {{Martinez}}}]{fermi11dsph}%
  \BibitemOpen
  \bibfield  {author} {\bibinfo {author} {\bibfnamefont {M.}~\bibnamefont
  {{Ackermann}}} {\it et~al.},\ }\bibfield  {title} {\enquote {\bibinfo {title}
  {{Constraining Dark Matter Models from a Combined Analysis of Milky Way
  Satellites with the Fermi Large Area Telescope}},}\ }\href {\doibase
  10.1103/PhysRevLett.107.241302} {\bibfield  {journal} {\bibinfo  {journal}
  {Physical Review Letters}\ }\textbf {\bibinfo {volume} {107}},\ \bibinfo
  {eid} {241302} (\bibinfo {year} {2011})},\ \Eprint
  {http://arxiv.org/abs/1108.3546}{arXiv:1108.3546}\BibitemShut {NoStop}%
\bibitem [{\citenamefont {Geringer-Sameth}\ and\ \citenamefont
  {Koushiappas}(2011)}]{GeringerSameth:2011iw}%
  \BibitemOpen
  \bibfield  {author} {\bibinfo {author} {\bibfnamefont {A.}~\bibnamefont
  {Geringer-Sameth}} and\ \bibinfo {author} {\bibfnamefont {S.~M.}\
  \bibnamefont {Koushiappas}},\ }\bibfield  {title} {\enquote {\bibinfo {title}
  {{Exclusion of canonical WIMPs by the joint analysis of Milky Way dwarfs with
  Fermi}},}\ }\href {\doibase 10.1103/PhysRevLett.107.241303} {\bibfield
  {journal} {\bibinfo  {journal} {Phys. Rev. Lett.}\ }\textbf {\bibinfo
  {volume} {107}},\ \bibinfo {pages} {241303} (\bibinfo {year} {2011})},\
  \Eprint {http://arxiv.org/abs/1108.2914}{arXiv:1108.2914}\BibitemShut
  {NoStop}%
\bibitem [{\citenamefont {{Cholis}}\ and\ \citenamefont
  {{Salucci}}(2012)}]{2012PhRvD86b3528C}%
  \BibitemOpen
  \bibfield  {author} {\bibinfo {author} {\bibfnamefont {I.}~\bibnamefont
  {{Cholis}}} and\ \bibinfo {author} {\bibfnamefont {P.}~\bibnamefont
  {{Salucci}}},\ }\bibfield  {title} {\enquote {\bibinfo {title} {{Extracting
  limits on dark matter annihilation from gamma ray observations towards dwarf
  spheroidal galaxies}},}\ }\href {\doibase 10.1103/PhysRevD.86.023528}
  {\bibfield  {journal} {\bibinfo  {journal} {\prd}\ }\textbf {\bibinfo
  {volume} {86}},\ \bibinfo {eid} {023528} (\bibinfo {year} {2012})},\ \Eprint
  {http://arxiv.org/abs/1203.2954}{arXiv:1203.2954}\BibitemShut {NoStop}%
\bibitem [{\citenamefont {Tsai}\ \emph {{\it et~al.}}(2013)\citenamefont
  {Tsai}, \citenamefont {Yuan},\ and\ \citenamefont {Huang}}]{tsai13dsph}%
  \BibitemOpen
  \bibfield  {author} {\bibinfo {author} {\bibfnamefont {Y.-L.~S.}\
  \bibnamefont {Tsai}}, \bibinfo {author} {\bibfnamefont {Q.}~\bibnamefont
  {Yuan}}, and\ \bibinfo {author} {\bibfnamefont {X.}~\bibnamefont {Huang}},\
  }\bibfield  {title} {\enquote {\bibinfo {title} {{A generic method to
  constrain the dark matter model parameters from Fermi observations of dwarf
  spheroids}},}\ }\href {\doibase 10.1088/1475-7516/2013/03/018} {\bibfield
  {journal} {\bibinfo  {journal} {JCAP}\ }\textbf {\bibinfo {volume} {1303}},\
  \bibinfo {pages} {018} (\bibinfo {year} {2013})},\ \Eprint
  {http://arxiv.org/abs/1212.3990}{arXiv:1212.3990}\BibitemShut {NoStop}%
\bibitem [{\citenamefont {{Ackermann}}\ \emph {{\it et~al.}}(2014)\citenamefont
  {{Ackermann}}, \citenamefont {{Albert}}, \citenamefont {{Anderson}},
  \citenamefont {{Baldini}}, \citenamefont {{Ballet}}, \citenamefont
  {{Barbiellini}}, \citenamefont {{Bastieri}}, \citenamefont {{Bechtol}},
  \citenamefont {{Bellazzini}}, \citenamefont {{Bissaldi}}, \citenamefont
  {{Bloom}}, \citenamefont {{Bonamente}}, \citenamefont {{Bouvier}},
  \citenamefont {{Brandt}}, \citenamefont {{Bregeon}}, \citenamefont
  {{Brigida}}, \citenamefont {{Bruel}}, \citenamefont {{Buehler}},
  \citenamefont {{Buson}}, \citenamefont {{Caliandro}}, \citenamefont
  {{Cameron}}, \citenamefont {{Caragiulo}}, \citenamefont {{Caraveo}},
  \citenamefont {{Cecchi}}, \citenamefont {{Charles}}, \citenamefont
  {{Chekhtman}}, \citenamefont {{Chiang}}, \citenamefont {{Ciprini}},
  \citenamefont {{Claus}}, \citenamefont {{Cohen-Tanugi}}, \citenamefont
  {{Conrad}}, \citenamefont {{D'Ammando}}, \citenamefont {{de Angelis}},
  \citenamefont {{Dermer}}, \citenamefont {{Digel}}, \citenamefont {{do Couto e
  Silva}}, \citenamefont {{Drell}}, \citenamefont {{Drlica-Wagner}},
  \citenamefont {{Essig}}, \citenamefont {{Favuzzi}}, \citenamefont
  {{Ferrara}}, \citenamefont {{Franckowiak}}, \citenamefont {{Fukazawa}},
  \citenamefont {{Funk}}, \citenamefont {{Fusco}}, \citenamefont {{Gargano}},
  \citenamefont {{Gasparrini}}, \citenamefont {{Giglietto}}, \citenamefont
  {{Giroletti}}, \citenamefont {{Godfrey}}, \citenamefont {{Gomez-Vargas}},
  \citenamefont {{Grenier}}, \citenamefont {{Guiriec}}, \citenamefont
  {{Gustafsson}}, \citenamefont {{Hayashida}}, \citenamefont {{Hays}},
  \citenamefont {{Hewitt}}, \citenamefont {{Hughes}}, \citenamefont {{Jogler}},
  \citenamefont {{Kamae}}, \citenamefont {{Kn{\"o}dlseder}}, \citenamefont
  {{Kocevski}}, \citenamefont {{Kuss}}, \citenamefont {{Larsson}},
  \citenamefont {{Latronico}}, \citenamefont {{Llena Garde}}, \citenamefont
  {{Longo}}, \citenamefont {{Loparco}}, \citenamefont {{Lovellette}},
  \citenamefont {{Lubrano}}, \citenamefont {{Martinez}}, \citenamefont
  {{Mayer}}, \citenamefont {{Mazziotta}}, \citenamefont {{Michelson}},
  \citenamefont {{Mitthumsiri}}, \citenamefont {{Mizuno}}, \citenamefont
  {{Moiseev}}, \citenamefont {{Monzani}}, \citenamefont {{Morselli}},
  \citenamefont {{Moskalenko}}, \citenamefont {{Murgia}}, \citenamefont
  {{Nemmen}}, \citenamefont {{Nuss}}, \citenamefont {{Ohsugi}}, \citenamefont
  {{Orlando}}, \citenamefont {{Ormes}}, \citenamefont {{Perkins}},
  \citenamefont {{Piron}}, \citenamefont {{Pivato}}, \citenamefont {{Porter}},
  \citenamefont {{Rain{\`o}}}, \citenamefont {{Rando}}, \citenamefont
  {{Razzano}}, \citenamefont {{Razzaque}}, \citenamefont {{Reimer}},
  \citenamefont {{Reimer}}, \citenamefont {{Ritz}}, \citenamefont
  {{S{\'a}nchez-Conde}}, \citenamefont {{Sehgal}}, \citenamefont {{Sgr{\`o}}},
  \citenamefont {{Siskind}}, \citenamefont {{Spinelli}}, \citenamefont
  {{Strigari}}, \citenamefont {{Suson}}, \citenamefont {{Tajima}},
  \citenamefont {{Takahashi}}, \citenamefont {{Thayer}}, \citenamefont
  {{Tibaldo}}, \citenamefont {{Tinivella}}, \citenamefont {{Torres}},
  \citenamefont {{Uchiyama}}, \citenamefont {{Usher}}, \citenamefont
  {{Vandenbroucke}}, \citenamefont {{Vianello}}, \citenamefont {{Vitale}},
  \citenamefont {{Werner}}, \citenamefont {{Winer}}, \citenamefont {{Wood}},
  \citenamefont {{Wood}}, \citenamefont {{Zaharijas}}, \citenamefont
  {{Zimmer}},\ and\ \citenamefont {{Fermi-LAT Collaboration}}}]{fermi14dsph}%
  \BibitemOpen
  \bibfield  {author} {\bibinfo {author} {\bibfnamefont {M.}~\bibnamefont
  {{Ackermann}}} {\it et~al.},\ }\bibfield  {title} {\enquote {\bibinfo {title}
  {{Dark matter constraints from observations of 25 Milky{\^A} Way satellite
  galaxies with the Fermi Large Area Telescope}},}\ }\href {\doibase
  10.1103/PhysRevD.89.042001} {\bibfield  {journal} {\bibinfo  {journal}
  {\prd}\ }\textbf {\bibinfo {volume} {89}},\ \bibinfo {eid} {042001} (\bibinfo
  {year} {2014})},\ \Eprint
  {http://arxiv.org/abs/1310.0828}{arXiv:1310.0828}\BibitemShut {NoStop}%
\bibitem [{\citenamefont {Zhao}\ \emph {{\it et~al.}}(2016)\citenamefont
  {Zhao}, \citenamefont {Bi}, \citenamefont {Jia}, \citenamefont {Yin},\ and\
  \citenamefont {Zhu}}]{zhao2016ds}%
  \BibitemOpen
  \bibfield  {author} {\bibinfo {author} {\bibfnamefont {Y.}~\bibnamefont
  {Zhao}}, \bibinfo {author} {\bibfnamefont {X.-J.}\ \bibnamefont {Bi}},
  \bibinfo {author} {\bibfnamefont {H.-Y.}\ \bibnamefont {Jia}}, \bibinfo
  {author} {\bibfnamefont {P.-F.}\ \bibnamefont {Yin}}, and\ \bibinfo {author}
  {\bibfnamefont {F.-R.}\ \bibnamefont {Zhu}},\ }\bibfield  {title} {\enquote
  {\bibinfo {title} {{Constraint on the velocity dependent dark matter
  annihilation cross section from Fermi-LAT observations of dwarf galaxies}},}\
  }\href {\doibase 10.1103/PhysRevD.93.083513} {\bibfield  {journal} {\bibinfo
  {journal} {Phys. Rev.}\ }\textbf {\bibinfo {volume} {D93}},\ \bibinfo {pages}
  {083513} (\bibinfo {year} {2016})},\ \Eprint
  {http://arxiv.org/abs/1601.02181}{arXiv:1601.02181}\BibitemShut {NoStop}%
\bibitem [{\citenamefont {{Geringer-Sameth}}\ \emph {{\it
  et~al.}}(2015)\citenamefont {{Geringer-Sameth}}, \citenamefont
  {{Koushiappas}},\ and\ \citenamefont {{Walker}}}]{gs15dsph}%
  \BibitemOpen
  \bibfield  {author} {\bibinfo {author} {\bibfnamefont {A.}~\bibnamefont
  {{Geringer-Sameth}}}, \bibinfo {author} {\bibfnamefont {S.~M.}\ \bibnamefont
  {{Koushiappas}}}, and\ \bibinfo {author} {\bibfnamefont {M.~G.}\ \bibnamefont
  {{Walker}}},\ }\bibfield  {title} {\enquote {\bibinfo {title} {{Comprehensive
  search for dark matter annihilation in dwarf galaxies}},}\ }\href {\doibase
  10.1103/PhysRevD.91.083535} {\bibfield  {journal} {\bibinfo  {journal}
  {\prd}\ }\textbf {\bibinfo {volume} {91}},\ \bibinfo {eid} {083535} (\bibinfo
  {year} {2015})},\ \Eprint
  {http://arxiv.org/abs/1410.2242}{arXiv:1410.2242}\BibitemShut {NoStop}%
\bibitem [{\citenamefont {{Ackermann}}\ \emph {{\it et~al.}}(2015)\citenamefont
  {{Ackermann}}, \citenamefont {{Albert}}, \citenamefont {{Anderson}},
  \citenamefont {{Atwood}}, \citenamefont {{Baldini}}, \citenamefont
  {{Barbiellini}}, \citenamefont {{Bastieri}}, \citenamefont {{Bechtol}},
  \citenamefont {{Bellazzini}}, \citenamefont {{Bissaldi}}, \citenamefont
  {{Blandford}}, \citenamefont {{Bloom}}, \citenamefont {{Bonino}},
  \citenamefont {{Bottacini}}, \citenamefont {{Brandt}}, \citenamefont
  {{Bregeon}}, \citenamefont {{Bruel}}, \citenamefont {{Buehler}},
  \citenamefont {{Caliandro}}, \citenamefont {{Cameron}}, \citenamefont
  {{Caputo}}, \citenamefont {{Caragiulo}}, \citenamefont {{Caraveo}},
  \citenamefont {{Cecchi}}, \citenamefont {{Charles}}, \citenamefont
  {{Chekhtman}}, \citenamefont {{Chiang}}, \citenamefont {{Chiaro}},
  \citenamefont {{Ciprini}}, \citenamefont {{Claus}}, \citenamefont
  {{Cohen-Tanugi}}, \citenamefont {{Conrad}}, \citenamefont {{Cuoco}},
  \citenamefont {{Cutini}}, \citenamefont {{D'Ammando}}, \citenamefont {{de
  Angelis}}, \citenamefont {{de Palma}}, \citenamefont {{Desiante}},
  \citenamefont {{Digel}}, \citenamefont {{Di Venere}}, \citenamefont
  {{Drell}}, \citenamefont {{Drlica-Wagner}}, \citenamefont {{Essig}},
  \citenamefont {{Favuzzi}}, \citenamefont {{Fegan}}, \citenamefont
  {{Ferrara}}, \citenamefont {{Focke}}, \citenamefont {{Franckowiak}},
  \citenamefont {{Fukazawa}}, \citenamefont {{Funk}}, \citenamefont {{Fusco}},
  \citenamefont {{Gargano}}, \citenamefont {{Gasparrini}}, \citenamefont
  {{Giglietto}}, \citenamefont {{Giordano}}, \citenamefont {{Giroletti}},
  \citenamefont {{Glanzman}}, \citenamefont {{Godfrey}}, \citenamefont
  {{Gomez-Vargas}}, \citenamefont {{Grenier}}, \citenamefont {{Guiriec}},
  \citenamefont {{Gustafsson}}, \citenamefont {{Hays}}, \citenamefont
  {{Hewitt}}, \citenamefont {{Horan}}, \citenamefont {{Jogler}}, \citenamefont
  {{J{\'o}hannesson}}, \citenamefont {{Kuss}}, \citenamefont {{Larsson}},
  \citenamefont {{Latronico}}, \citenamefont {{Li}}, \citenamefont {{Li}},
  \citenamefont {{Llena Garde}}, \citenamefont {{Longo}}, \citenamefont
  {{Loparco}}, \citenamefont {{Lubrano}}, \citenamefont {{Malyshev}},
  \citenamefont {{Mayer}}, \citenamefont {{Mazziotta}}, \citenamefont
  {{McEnery}}, \citenamefont {{Meyer}}, \citenamefont {{Michelson}},
  \citenamefont {{Mizuno}}, \citenamefont {{Moiseev}}, \citenamefont
  {{Monzani}}, \citenamefont {{Morselli}}, \citenamefont {{Murgia}},
  \citenamefont {{Nuss}}, \citenamefont {{Ohsugi}}, \citenamefont {{Orienti}},
  \citenamefont {{Orlando}}, \citenamefont {{Ormes}}, \citenamefont
  {{Paneque}}, \citenamefont {{Perkins}}, \citenamefont {{Pesce-Rollins}},
  \citenamefont {{Piron}}, \citenamefont {{Pivato}}, \citenamefont {{Porter}},
  \citenamefont {{Rain{\`o}}}, \citenamefont {{Rando}}, \citenamefont
  {{Razzano}}, \citenamefont {{Reimer}}, \citenamefont {{Reimer}},
  \citenamefont {{Ritz}}, \citenamefont {{S{\'a}nchez-Conde}}, \citenamefont
  {{Schulz}}, \citenamefont {{Sehgal}}, \citenamefont {{Sgr{\`o}}},
  \citenamefont {{Siskind}}, \citenamefont {{Spada}}, \citenamefont
  {{Spandre}}, \citenamefont {{Spinelli}}, \citenamefont {{Strigari}},
  \citenamefont {{Tajima}}, \citenamefont {{Takahashi}}, \citenamefont
  {{Thayer}}, \citenamefont {{Tibaldo}}, \citenamefont {{Torres}},
  \citenamefont {{Troja}}, \citenamefont {{Vianello}}, \citenamefont
  {{Werner}}, \citenamefont {{Winer}}, \citenamefont {{Wood}}, \citenamefont
  {{Wood}}, \citenamefont {{Zaharijas}}, \citenamefont {{Zimmer}},\ and\
  \citenamefont {{Fermi-LAT Collaboration}}}]{fermi15dsph}%
  \BibitemOpen
  \bibfield  {author} {\bibinfo {author} {\bibfnamefont {M.}~\bibnamefont
  {{Ackermann}}} {\it et~al.},\ }\bibfield  {title} {\enquote {\bibinfo {title}
  {{Searching for Dark Matter Annihilation from Milky Way Dwarf Spheroidal
  Galaxies with Six Years of Fermi Large Area Telescope Data}},}\ }\href
  {\doibase 10.1103/PhysRevLett.115.231301} {\bibfield  {journal} {\bibinfo
  {journal} {Physical Review Letters}\ }\textbf {\bibinfo {volume} {115}},\
  \bibinfo {eid} {231301} (\bibinfo {year} {2015})},\ \Eprint
  {http://arxiv.org/abs/1503.02641}{arXiv:1503.02641}\BibitemShut {NoStop}%
\bibitem [{\citenamefont {Simon}\ \emph {{\it et~al.}}(2015)\citenamefont
  {Simon} {\it et~al.}}]{Simon:2015fdw}%
  \BibitemOpen
  \bibfield  {author} {\bibinfo {author} {\bibfnamefont {J.~D.}\ \bibnamefont
  {Simon}}  {\it et~al.} (\bibinfo {collaboration} {DES} Collaboration),\
  }\bibfield  {title} {\enquote {\bibinfo {title} {{Stellar Kinematics and
  Metallicities in the Ultra-Faint Dwarf Galaxy Reticulum II}},}\ }\href
  {\doibase 10.1088/0004-637X/808/1/95} {\bibfield  {journal} {\bibinfo
  {journal} {Astrophys. J.}\ }\textbf {\bibinfo {volume} {808}},\ \bibinfo
  {pages} {95} (\bibinfo {year} {2015})},\ \Eprint
  {http://arxiv.org/abs/1504.02889}{arXiv:1504.02889}\BibitemShut {NoStop}%
\bibitem [{\citenamefont {Geringer-Sameth}\ \emph {{\it
  et~al.}}(2015{\natexlab{a}})\citenamefont {Geringer-Sameth}, \citenamefont
  {Koushiappas},\ and\ \citenamefont {Walker}}]{Geringer-Sameth:2014yza}%
  \BibitemOpen
  \bibfield  {author} {\bibinfo {author} {\bibfnamefont {A.}~\bibnamefont
  {Geringer-Sameth}}, \bibinfo {author} {\bibfnamefont {S.~M.}\ \bibnamefont
  {Koushiappas}}, and\ \bibinfo {author} {\bibfnamefont {M.}~\bibnamefont
  {Walker}},\ }\bibfield  {title} {\enquote {\bibinfo {title} {{Dwarf galaxy
  annihilation and decay emission profiles for dark matter experiments}},}\
  }\href {\doibase 10.1088/0004-637X/801/2/74} {\bibfield  {journal} {\bibinfo
  {journal} {Astrophys. J.}\ }\textbf {\bibinfo {volume} {801}},\ \bibinfo
  {pages} {74} (\bibinfo {year} {2015}{\natexlab{a}})},\ \Eprint
  {http://arxiv.org/abs/1408.0002}{arXiv:1408.0002}\BibitemShut {NoStop}%
\bibitem [{\citenamefont {Drlica-Wagner}\ \emph {{\it
  et~al.}}(2015{\natexlab{a}})\citenamefont {Drlica-Wagner} {\it
  et~al.}}]{Drlica-Wagner:2015xua}%
  \BibitemOpen
  \bibfield  {author} {\bibinfo {author} {\bibfnamefont {A.}~\bibnamefont
  {Drlica-Wagner}}  {\it et~al.} (\bibinfo {collaboration} {DES, Fermi-LAT}
  Collaboration),\ }\bibfield  {title} {\enquote {\bibinfo {title} {{Search for
  Gamma-Ray Emission from DES Dwarf Spheroidal Galaxy Candidates with Fermi-LAT
  Data}},}\ }\href {\doibase 10.1088/2041-8205/809/1/L4} {\bibfield  {journal}
  {\bibinfo  {journal} {Astrophys. J.}\ }\textbf {\bibinfo {volume} {809}},\
  \bibinfo {pages} {L4} (\bibinfo {year} {2015}{\natexlab{a}})},\ \Eprint
  {http://arxiv.org/abs/1503.02632}{arXiv:1503.02632}\BibitemShut {NoStop}%
\bibitem [{\citenamefont {Bechtol}\ \emph {{\it et~al.}}(2015)\citenamefont
  {Bechtol} {\it et~al.}}]{des15y1}%
  \BibitemOpen
  \bibfield  {author} {\bibinfo {author} {\bibfnamefont {K.}~\bibnamefont
  {Bechtol}}  {\it et~al.} (\bibinfo {collaboration} {DES} Collaboration),\
  }\bibfield  {title} {\enquote {\bibinfo {title} {{Eight New Milky Way
  Companions Discovered in First-Year Dark Energy Survey Data}},}\ }\href
  {\doibase 10.1088/0004-637X/807/1/50} {\bibfield  {journal} {\bibinfo
  {journal} {Astrophys. J.}\ }\textbf {\bibinfo {volume} {807}},\ \bibinfo
  {pages} {50} (\bibinfo {year} {2015})},\ \Eprint
  {http://arxiv.org/abs/1503.02584}{arXiv:1503.02584}\BibitemShut {NoStop}%
\bibitem [{\citenamefont {Drlica-Wagner}\ \emph {{\it
  et~al.}}(2015{\natexlab{b}})\citenamefont {Drlica-Wagner} {\it
  et~al.}}]{des15y2}%
  \BibitemOpen
  \bibfield  {author} {\bibinfo {author} {\bibfnamefont {A.}~\bibnamefont
  {Drlica-Wagner}}  {\it et~al.} (\bibinfo {collaboration} {DES}
  Collaboration),\ }\bibfield  {title} {\enquote {\bibinfo {title} {{Eight
  Ultra-faint Galaxy Candidates Discovered in Year Two of the Dark Energy
  Survey}},}\ }\href {\doibase 10.1088/0004-637X/813/2/109} {\bibfield
  {journal} {\bibinfo  {journal} {Astrophys. J.}\ }\textbf {\bibinfo {volume}
  {813}},\ \bibinfo {pages} {109} (\bibinfo {year} {2015}{\natexlab{b}})},\
  \Eprint {http://arxiv.org/abs/1508.03622}{arXiv:1508.03622}\BibitemShut
  {NoStop}%
\bibitem [{\citenamefont {Laevens}\ \emph {{\it
  et~al.}}(2015{\natexlab{a}})\citenamefont {Laevens} {\it
  et~al.}}]{laevens15tri2}%
  \BibitemOpen
  \bibfield  {author} {\bibinfo {author} {\bibfnamefont {B.~P.~M.}\
  \bibnamefont {Laevens}}  {\it et~al.},\ }\bibfield  {title} {\enquote
  {\bibinfo {title} {{A New Faint Milky Way Satellite Discovered in the
  Pan-STARRS1 3 pi Survey}},}\ }\href {\doibase 10.1088/2041-8205/802/2/L18}
  {\bibfield  {journal} {\bibinfo  {journal} {Astrophys. J.}\ }\textbf
  {\bibinfo {volume} {802}},\ \bibinfo {pages} {L18} (\bibinfo {year}
  {2015}{\natexlab{a}})},\ \Eprint
  {http://arxiv.org/abs/1503.05554}{arXiv:1503.05554}\BibitemShut {NoStop}%
\bibitem [{\citenamefont {Laevens}\ \emph {{\it
  et~al.}}(2015{\natexlab{b}})\citenamefont {Laevens} {\it
  et~al.}}]{laevens15_3}%
  \BibitemOpen
  \bibfield  {author} {\bibinfo {author} {\bibfnamefont {B.~P.~M.}\
  \bibnamefont {Laevens}}  {\it et~al.},\ }\bibfield  {title} {\enquote
  {\bibinfo {title} {{Sagittarius II, Draco II and Laevens 3: Three new Milky
  way Satellites Discovered in the Pan-starrs 1 3$\pi$ Survey}},}\ }\href
  {\doibase 10.1088/0004-637X/813/1/44} {\bibfield  {journal} {\bibinfo
  {journal} {Astrophys. J.}\ }\textbf {\bibinfo {volume} {813}},\ \bibinfo
  {pages} {44} (\bibinfo {year} {2015}{\natexlab{b}})},\ \Eprint
  {http://arxiv.org/abs/1507.07564}{arXiv:1507.07564}\BibitemShut {NoStop}%
\bibitem [{\citenamefont {Kim}\ \emph {{\it et~al.}}(2015)\citenamefont {Kim},
  \citenamefont {Jerjen}, \citenamefont {Mackey}, \citenamefont {Da~Costa},\
  and\ \citenamefont {Milone}}]{kim15pegasus3}%
  \BibitemOpen
  \bibfield  {author} {\bibinfo {author} {\bibfnamefont {D.}~\bibnamefont
  {Kim}}, \bibinfo {author} {\bibfnamefont {H.}~\bibnamefont {Jerjen}},
  \bibinfo {author} {\bibfnamefont {D.}~\bibnamefont {Mackey}}, \bibinfo
  {author} {\bibfnamefont {G.~S.}\ \bibnamefont {Da~Costa}}, and\ \bibinfo
  {author} {\bibfnamefont {A.~P.}\ \bibnamefont {Milone}},\ }\bibfield  {title}
  {\enquote {\bibinfo {title} {{A Hero’s Dark Horse: Discovery of an
  Ultra-faint Milky way Satellite in Pegasus}},}\ }\href {\doibase
  10.1088/2041-8205/804/2/L44} {\bibfield  {journal} {\bibinfo  {journal}
  {Astrophys. J.}\ }\textbf {\bibinfo {volume} {804}},\ \bibinfo {pages} {L44}
  (\bibinfo {year} {2015})},\ \Eprint
  {http://arxiv.org/abs/1503.08268}{arXiv:1503.08268}\BibitemShut {NoStop}%
\bibitem [{\citenamefont {Homma}\ \emph {{\it et~al.}}(2016)\citenamefont
  {Homma} {\it et~al.}}]{Daisuke(2016)}%
  \BibitemOpen
  \bibfield  {author} {\bibinfo {author} {\bibfnamefont {D.}~\bibnamefont
  {Homma}}  {\it et~al.},\ }\bibfield  {title} {\enquote {\bibinfo {title} {{A
  New Milky Way Satellite Discovered In The Subaru/Hyper Suprime-Cam
  Survey}},}\ }\href {\doibase 10.3847/0004-637X/832/1/21} {\bibfield
  {journal} {\bibinfo  {journal} {Astrophys. J.}\ }\textbf {\bibinfo {volume}
  {832}},\ \bibinfo {pages} {21} (\bibinfo {year} {2016})},\ \Eprint
  {http://arxiv.org/abs/1609.04346}{arXiv:1609.04346}\BibitemShut {NoStop}%
\bibitem [{\citenamefont {{Homma}}\ \emph {{\it et~al.}}(2017)\citenamefont
  {{Homma}}, \citenamefont {{Chiba}}, \citenamefont {{Okamoto}}, \citenamefont
  {{Komiyama}}, \citenamefont {{Tanaka}}, \citenamefont {{Tanaka}},
  \citenamefont {{Ishigaki}}, \citenamefont {{Hayashi}}, \citenamefont
  {{Arimoto}}, \citenamefont {{Garmilla}}, \citenamefont {{Lupton}},
  \citenamefont {{Strauss}}, \citenamefont {{Miyazaki}}, \citenamefont
  {{Wang}},\ and\ \citenamefont {{Murayama}}}]{Daisuke(2017)}%
  \BibitemOpen
  \bibfield  {author} {\bibinfo {author} {\bibfnamefont {D.}~\bibnamefont
  {{Homma}}} {\it et~al.},\ }\bibfield  {title} {\enquote {\bibinfo {title}
  {{Searches for New Milky Way Satellites from the First Two Years of Data of
  the Subaru/Hyper Suprime-Cam Survey: Discovery of Cetus\~{}III}},}\
  }\href@noop {} {\bibfield  {journal} {\bibinfo  {journal} {ArXiv e-prints}\ }
  (\bibinfo {year} {2017})},\ \Eprint
  {http://arxiv.org/abs/1704.05977}{arXiv:1704.05977}\BibitemShut {NoStop}%
\bibitem [{\citenamefont {Drlica-Wagner}\ \emph {{\it
  et~al.}}(2016)\citenamefont {Drlica-Wagner} {\it
  et~al.}}]{Drlica-Wagner1(2016)}%
  \BibitemOpen
  \bibfield  {author} {\bibinfo {author} {\bibfnamefont {A.}~\bibnamefont
  {Drlica-Wagner}}  {\it et~al.},\ }\bibfield  {title} {\enquote {\bibinfo
  {title} {{An Ultra-Faint Galaxy Candidate Discovered in Early Data from the
  Magellanic Satellites Survey}},}\ }\href {\doibase
  10.3847/2041-8205/833/1/L5} {\bibfield  {journal} {\bibinfo  {journal}
  {Astrophys. J.}\ }\textbf {\bibinfo {volume} {833}},\ \bibinfo {pages} {L5}
  (\bibinfo {year} {2016})},\ \Eprint
  {http://arxiv.org/abs/1609.02148}{arXiv:1609.02148}\BibitemShut {NoStop}%
\bibitem [{\citenamefont {Torrealba}\ \emph {{\it et~al.}}(2016)\citenamefont
  {Torrealba}, \citenamefont {Koposov}, \citenamefont {Belokurov},
  \citenamefont {Irwin}, \citenamefont {Collins}, \citenamefont {Spencer},
  \citenamefont {Ibata}, \citenamefont {Mateo}, \citenamefont {Bonaca},\ and\
  \citenamefont {Jethwa}}]{Torrealba(2016)}%
  \BibitemOpen
  \bibfield  {author} {\bibinfo {author} {\bibfnamefont {G.}~\bibnamefont
  {Torrealba}}, \bibinfo {author} {\bibfnamefont {S.~E.}\ \bibnamefont
  {Koposov}}, \bibinfo {author} {\bibfnamefont {V.}~\bibnamefont {Belokurov}},
  \bibinfo {author} {\bibfnamefont {M.}~\bibnamefont {Irwin}}, \bibinfo
  {author} {\bibfnamefont {M.}~\bibnamefont {Collins}}, \bibinfo {author}
  {\bibfnamefont {M.}~\bibnamefont {Spencer}}, \bibinfo {author} {\bibfnamefont
  {R.}~\bibnamefont {Ibata}}, \bibinfo {author} {\bibfnamefont
  {M.}~\bibnamefont {Mateo}}, \bibinfo {author} {\bibfnamefont
  {A.}~\bibnamefont {Bonaca}}, and\ \bibinfo {author} {\bibfnamefont
  {P.}~\bibnamefont {Jethwa}},\ }\bibfield  {title} {\enquote {\bibinfo {title}
  {{At the survey limits: discovery of the Aquarius 2 dwarf galaxy in the VST
  ATLAS and the SDSS data}},}\ }\href {\doibase 10.1093/mnras/stw2051}
  {\bibfield  {journal} {\bibinfo  {journal} {Monthly Notices of the Royal
  Astronomical Society}\ }\textbf {\bibinfo {volume} {463}},\ \bibinfo {pages}
  {712} (\bibinfo {year} {2016})},\ \Eprint
  {http://arxiv.org/abs/1605.05338}{arXiv:1605.05338}\BibitemShut {NoStop}%
\bibitem [{\citenamefont {Geringer-Sameth}\ \emph {{\it
  et~al.}}(2015{\natexlab{b}})\citenamefont {Geringer-Sameth}, \citenamefont
  {Walker}, \citenamefont {Koushiappas}, \citenamefont {Koposov}, \citenamefont
  {Belokurov}, \citenamefont {Torrealba},\ and\ \citenamefont
  {Evans}}]{gs15ret2}%
  \BibitemOpen
  \bibfield  {author} {\bibinfo {author} {\bibfnamefont {A.}~\bibnamefont
  {Geringer-Sameth}}, \bibinfo {author} {\bibfnamefont {M.~G.}\ \bibnamefont
  {Walker}}, \bibinfo {author} {\bibfnamefont {S.~M.}\ \bibnamefont
  {Koushiappas}}, \bibinfo {author} {\bibfnamefont {S.~E.}\ \bibnamefont
  {Koposov}}, \bibinfo {author} {\bibfnamefont {V.}~\bibnamefont {Belokurov}},
  \bibinfo {author} {\bibfnamefont {G.}~\bibnamefont {Torrealba}}, and\
  \bibinfo {author} {\bibfnamefont {N.~W.}\ \bibnamefont {Evans}},\ }\bibfield
  {title} {\enquote {\bibinfo {title} {{Indication of Gamma-ray Emission from
  the Newly Discovered Dwarf Galaxy Reticulum II}},}\ }\href {\doibase
  10.1103/PhysRevLett.115.081101} {\bibfield  {journal} {\bibinfo  {journal}
  {Phys. Rev. Lett.}\ }\textbf {\bibinfo {volume} {115}},\ \bibinfo {pages}
  {081101} (\bibinfo {year} {2015}{\natexlab{b}})},\ \Eprint
  {http://arxiv.org/abs/1503.02320}{arXiv:1503.02320}\BibitemShut {NoStop}%
\bibitem [{\citenamefont {Hooper}\ and\ \citenamefont
  {Linden}(2015)}]{hooper15ret2}%
  \BibitemOpen
  \bibfield  {author} {\bibinfo {author} {\bibfnamefont {D.}~\bibnamefont
  {Hooper}} and\ \bibinfo {author} {\bibfnamefont {T.}~\bibnamefont {Linden}},\
  }\bibfield  {title} {\enquote {\bibinfo {title} {{On The Gamma-Ray Emission
  From Reticulum II and Other Dwarf Galaxies}},}\ }\href {\doibase
  10.1088/1475-7516/2015/09/016} {\bibfield  {journal} {\bibinfo  {journal}
  {JCAP}\ }\textbf {\bibinfo {volume} {1509}},\ \bibinfo {pages} {016}
  (\bibinfo {year} {2015})},\ \Eprint
  {http://arxiv.org/abs/1503.06209}{arXiv:1503.06209}\BibitemShut {NoStop}%
\bibitem [{\citenamefont {Li}\ \emph {{\it et~al.}}(2016)\citenamefont {Li},
  \citenamefont {Liang}, \citenamefont {Duan}, \citenamefont {Shen},
  \citenamefont {Huang}, \citenamefont {Li}, \citenamefont {Fan}, \citenamefont
  {Liao}, \citenamefont {Feng},\ and\ \citenamefont {Chang}}]{li16dsph}%
  \BibitemOpen
  \bibfield  {author} {\bibinfo {author} {\bibfnamefont {S.}~\bibnamefont
  {Li}}, \bibinfo {author} {\bibfnamefont {Y.-F.}\ \bibnamefont {Liang}},
  \bibinfo {author} {\bibfnamefont {K.-K.}\ \bibnamefont {Duan}}, \bibinfo
  {author} {\bibfnamefont {Z.-Q.}\ \bibnamefont {Shen}}, \bibinfo {author}
  {\bibfnamefont {X.}~\bibnamefont {Huang}}, \bibinfo {author} {\bibfnamefont
  {X.}~\bibnamefont {Li}}, \bibinfo {author} {\bibfnamefont {Y.-Z.}\
  \bibnamefont {Fan}}, \bibinfo {author} {\bibfnamefont {N.-H.}\ \bibnamefont
  {Liao}}, \bibinfo {author} {\bibfnamefont {L.}~\bibnamefont {Feng}}, and\
  \bibinfo {author} {\bibfnamefont {J.}~\bibnamefont {Chang}},\ }\bibfield
  {title} {\enquote {\bibinfo {title} {{Search for gamma-ray emission from
  eight dwarf spheroidal galaxy candidates discovered in Year Two of Dark
  Energy Survey with Fermi-LAT data}},}\ }\href {\doibase
  10.1103/PhysRevD.93.043518} {\bibfield  {journal} {\bibinfo  {journal} {Phys.
  Rev. D}\ }\textbf {\bibinfo {volume} {93}},\ \bibinfo {pages} {043518}
  (\bibinfo {year} {2016})},\ \Eprint
  {http://arxiv.org/abs/1511.09252}{arXiv:1511.09252}\BibitemShut {NoStop}%
\bibitem [{\citenamefont {{Albert}}\ \emph {{\it et~al.}}(2017)\citenamefont
  {{Albert}}, \citenamefont {{Anderson}}, \citenamefont {{Bechtol}},
  \citenamefont {{Drlica-Wagner}}, \citenamefont {{Meyer}}, \citenamefont
  {{S{\'a}nchez-Conde}}, \citenamefont {{Strigari}}, \citenamefont {{Wood}},
  \citenamefont {{Abbott}}, \citenamefont {{Abdalla}}, \citenamefont
  {{Benoit-L{\'e}vy}}, \citenamefont {{Bernstein}}, \citenamefont
  {{Bernstein}}, \citenamefont {{Bertin}}, \citenamefont {{Brooks}},
  \citenamefont {{Burke}}, \citenamefont {{Carnero Rosell}}, \citenamefont
  {{Carrasco Kind}}, \citenamefont {{Carretero}}, \citenamefont {{Crocce}},
  \citenamefont {{Cunha}}, \citenamefont {{D'Andrea}}, \citenamefont {{da
  Costa}}, \citenamefont {{Desai}}, \citenamefont {{Diehl}}, \citenamefont
  {{Dietrich}}, \citenamefont {{Doel}}, \citenamefont {{Eifler}}, \citenamefont
  {{Evrard}}, \citenamefont {{Fausti Neto}}, \citenamefont {{Finley}},
  \citenamefont {{Flaugher}}, \citenamefont {{Fosalba}}, \citenamefont
  {{Frieman}}, \citenamefont {{Gerdes}}, \citenamefont {{Goldstein}},
  \citenamefont {{Gruen}}, \citenamefont {{Gruendl}}, \citenamefont
  {{Honscheid}}, \citenamefont {{James}}, \citenamefont {{Kent}}, \citenamefont
  {{Kuehn}}, \citenamefont {{Kuropatkin}}, \citenamefont {{Lahav}},
  \citenamefont {{Li}}, \citenamefont {{Maia}}, \citenamefont {{March}},
  \citenamefont {{Marshall}}, \citenamefont {{Martini}}, \citenamefont
  {{Miller}}, \citenamefont {{Miquel}}, \citenamefont {{Neilsen}},
  \citenamefont {{Nord}}, \citenamefont {{Ogando}}, \citenamefont {{Plazas}},
  \citenamefont {{Reil}}, \citenamefont {{Romer}}, \citenamefont {{Rykoff}},
  \citenamefont {{Sanchez}}, \citenamefont {{Santiago}}, \citenamefont
  {{Schubnell}}, \citenamefont {{Sevilla-Noarbe}}, \citenamefont {{Smith}},
  \citenamefont {{Soares-Santos}}, \citenamefont {{Sobreira}}, \citenamefont
  {{Suchyta}}, \citenamefont {{Swanson}}, \citenamefont {{Tarle}},
  \citenamefont {{Vikram}}, \citenamefont {{Walker}}, \citenamefont
  {{Wechsler}}, \citenamefont {{Fermi-LAT Collaboration}},\ and\ \citenamefont
  {{DES Collaboration}}}]{fermi2016dsph}%
  \BibitemOpen
  \bibfield  {author} {\bibinfo {author} {\bibfnamefont {A.}~\bibnamefont
  {{Albert}}} {\it et~al.},\ }\bibfield  {title} {\enquote {\bibinfo {title}
  {{Searching for Dark Matter Annihilation in Recently Discovered Milky Way
  Satellites with Fermi-Lat}},}\ }\href {\doibase 10.3847/1538-4357/834/2/110}
  {\bibfield  {journal} {\bibinfo  {journal} {Astrophys. J.}\ }\textbf
  {\bibinfo {volume} {834}},\ \bibinfo {eid} {110} (\bibinfo {year} {2017})},\
  \Eprint {http://arxiv.org/abs/1611.03184}{arXiv:1611.03184}\BibitemShut
  {NoStop}%
\bibitem [{\citenamefont {Liang}\ \emph {{\it et~al.}}(2016)\citenamefont
  {Liang}, \citenamefont {Xia}, \citenamefont {Shen}, \citenamefont {Li},
  \citenamefont {Jiang}, \citenamefont {Yuan}, \citenamefont {Fan},
  \citenamefont {Feng}, \citenamefont {Liang},\ and\ \citenamefont
  {Chang}}]{liang2016dl}%
  \BibitemOpen
  \bibfield  {author} {\bibinfo {author} {\bibfnamefont {Y.-F.}\ \bibnamefont
  {Liang}}, \bibinfo {author} {\bibfnamefont {Z.-Q.}\ \bibnamefont {Xia}},
  \bibinfo {author} {\bibfnamefont {Z.-Q.}\ \bibnamefont {Shen}}, \bibinfo
  {author} {\bibfnamefont {X.}~\bibnamefont {Li}}, \bibinfo {author}
  {\bibfnamefont {W.}~\bibnamefont {Jiang}}, \bibinfo {author} {\bibfnamefont
  {Q.}~\bibnamefont {Yuan}}, \bibinfo {author} {\bibfnamefont {Y.-Z.}\
  \bibnamefont {Fan}}, \bibinfo {author} {\bibfnamefont {L.}~\bibnamefont
  {Feng}}, \bibinfo {author} {\bibfnamefont {E.-W.}\ \bibnamefont {Liang}},
  and\ \bibinfo {author} {\bibfnamefont {J.}~\bibnamefont {Chang}},\ }\bibfield
   {title} {\enquote {\bibinfo {title} {{Search for gamma-ray line features
  from Milky Way satellites with Fermi LAT Pass 8 data}},}\ }\href {\doibase
  10.1103/PhysRevD.94.103502} {\bibfield  {journal} {\bibinfo  {journal} {Phys.
  Rev.}\ }\textbf {\bibinfo {volume} {D94}},\ \bibinfo {pages} {103502}
  (\bibinfo {year} {2016})},\ \Eprint
  {http://arxiv.org/abs/1608.07184}{arXiv:1608.07184}\BibitemShut {NoStop}%
\bibitem [{\citenamefont {{Li}}\ \emph {{\it et~al.}}(2018)\citenamefont
  {{Li}}, \citenamefont {{Duan}}, \citenamefont {{Liang}}, \citenamefont
  {{Xia}}, \citenamefont {{Shen}}, \citenamefont {{Li}}, \citenamefont
  {{Liao}}, \citenamefont {{Feng}}, \citenamefont {{Yuan}}, \citenamefont
  {{Fan}},\ and\ \citenamefont {{Chang}}}]{2018PhRvD..97l2001L}%
  \BibitemOpen
  \bibfield  {author} {\bibinfo {author} {\bibfnamefont {S.}~\bibnamefont
  {{Li}}} {\it et~al.},\ }\bibfield  {title} {\enquote {\bibinfo {title}
  {{Search for gamma-ray emission from the nearby dwarf spheroidal galaxies
  with 9 years of Fermi-LAT data}},}\ }\href {\doibase
  10.1103/PhysRevD.97.122001} {\bibfield  {journal} {\bibinfo  {journal}
  {\prd}\ }\textbf {\bibinfo {volume} {97}},\ \bibinfo {eid} {122001} (\bibinfo
  {year} {2018})},\ \Eprint
  {http://arxiv.org/abs/1805.06612}{arXiv:1805.06612}\BibitemShut {NoStop}%
\bibitem [{\citenamefont {Bhattacharjee}\ \emph {{\it
  et~al.}}(2019)\citenamefont {Bhattacharjee}, \citenamefont {Majumdar},
  \citenamefont {Biswas},\ and\ \citenamefont
  {Joarder}}]{Bhattacharjee:2018xem}%
  \BibitemOpen
  \bibfield  {author} {\bibinfo {author} {\bibfnamefont {P.}~\bibnamefont
  {Bhattacharjee}}, \bibinfo {author} {\bibfnamefont {P.}~\bibnamefont
  {Majumdar}}, \bibinfo {author} {\bibfnamefont {S.}~\bibnamefont {Biswas}},
  and\ \bibinfo {author} {\bibfnamefont {P.~S.}\ \bibnamefont {Joarder}},\
  }\bibfield  {title} {\enquote {\bibinfo {title} {{Analysis of Fermi-LAT data
  from Tucana-II: Possible constraints on the Dark Matter models with an
  intriguing hint of a signal}},}\ }\href {\doibase
  10.1088/1475-7516/2019/08/028} {\bibfield  {journal} {\bibinfo  {journal}
  {JCAP}\ }\textbf {\bibinfo {volume} {08}},\ \bibinfo {pages} {028} (\bibinfo
  {year} {2019})},\ \Eprint
  {http://arxiv.org/abs/1804.07542}{arXiv:1804.07542}\BibitemShut {NoStop}%
\bibitem [{\citenamefont {Hooper}\ and\ \citenamefont
  {Goodenough}(2011)}]{Hooper:2010mq}%
  \BibitemOpen
  \bibfield  {author} {\bibinfo {author} {\bibfnamefont {D.}~\bibnamefont
  {Hooper}} and\ \bibinfo {author} {\bibfnamefont {L.}~\bibnamefont
  {Goodenough}},\ }\bibfield  {title} {\enquote {\bibinfo {title} {{Dark Matter
  Annihilation in The Galactic Center As Seen by the Fermi Gamma Ray Space
  Telescope}},}\ }\href {\doibase 10.1016/j.physletb.2011.02.029} {\bibfield
  {journal} {\bibinfo  {journal} {Phys. Lett. B}\ }\textbf {\bibinfo {volume}
  {697}},\ \bibinfo {pages} {412} (\bibinfo {year} {2011})},\ \Eprint
  {http://arxiv.org/abs/1010.2752}{arXiv:1010.2752}\BibitemShut {NoStop}%
\bibitem [{\citenamefont {Gordon}\ and\ \citenamefont
  {Macias}(2013)}]{Gordon:2013vta}%
  \BibitemOpen
  \bibfield  {author} {\bibinfo {author} {\bibfnamefont {C.}~\bibnamefont
  {Gordon}} and\ \bibinfo {author} {\bibfnamefont {O.}~\bibnamefont {Macias}},\
  }\bibfield  {title} {\enquote {\bibinfo {title} {{Dark Matter and Pulsar
  Model Constraints from Galactic Center Fermi-LAT Gamma Ray Observations}},}\
  }\href {\doibase 10.1103/PhysRevD.88.083521, 10.1103/PhysRevD.89.049901}
  {\bibfield  {journal} {\bibinfo  {journal} {Phys. Rev. D}\ }\textbf {\bibinfo
  {volume} {88}},\ \bibinfo {pages} {083521} (\bibinfo {year} {2013})},\
  \bibinfo {note} {[Erratum: Phys. Rev.D89,no.4,049901(2014)]},\ \Eprint
  {http://arxiv.org/abs/1306.5725}{arXiv:1306.5725}\BibitemShut {NoStop}%
\bibitem [{\citenamefont {Hooper}\ and\ \citenamefont
  {Slatyer}(2013)}]{Hooper:2013rwa}%
  \BibitemOpen
  \bibfield  {author} {\bibinfo {author} {\bibfnamefont {D.}~\bibnamefont
  {Hooper}} and\ \bibinfo {author} {\bibfnamefont {T.~R.}\ \bibnamefont
  {Slatyer}},\ }\bibfield  {title} {\enquote {\bibinfo {title} {{Two Emission
  Mechanisms in the Fermi Bubbles: A Possible Signal of Annihilating Dark
  Matter}},}\ }\href {\doibase 10.1016/j.dark.2013.06.003} {\bibfield
  {journal} {\bibinfo  {journal} {Phys. Dark Univ.}\ }\textbf {\bibinfo
  {volume} {2}},\ \bibinfo {pages} {118} (\bibinfo {year} {2013})},\ \Eprint
  {http://arxiv.org/abs/1302.6589}{arXiv:1302.6589}\BibitemShut {NoStop}%
\bibitem [{\citenamefont {{Daylan}}\ \emph {{\it et~al.}}(2016)\citenamefont
  {{Daylan}}, \citenamefont {{Finkbeiner}}, \citenamefont {{Hooper}},
  \citenamefont {{Linden}}, \citenamefont {{Portillo}}, \citenamefont
  {{Rodd}},\ and\ \citenamefont {{Slatyer}}}]{Daylan:2014rsa}%
  \BibitemOpen
  \bibfield  {author} {\bibinfo {author} {\bibfnamefont {T.}~\bibnamefont
  {{Daylan}}}, \bibinfo {author} {\bibfnamefont {D.~P.}\ \bibnamefont
  {{Finkbeiner}}}, \bibinfo {author} {\bibfnamefont {D.}~\bibnamefont
  {{Hooper}}}, \bibinfo {author} {\bibfnamefont {T.}~\bibnamefont {{Linden}}},
  \bibinfo {author} {\bibfnamefont {S.~K.~N.}\ \bibnamefont {{Portillo}}},
  \bibinfo {author} {\bibfnamefont {N.~L.}\ \bibnamefont {{Rodd}}}, and\
  \bibinfo {author} {\bibfnamefont {T.~R.}\ \bibnamefont {{Slatyer}}},\
  }\bibfield  {title} {\enquote {\bibinfo {title} {{The characterization of the
  gamma-ray signal from the central Milky Way: A case for annihilating dark
  matter}},}\ }\href {\doibase 10.1016/j.dark.2015.12.005} {\bibfield
  {journal} {\bibinfo  {journal} {Physics of the Dark Universe}\ }\textbf
  {\bibinfo {volume} {12}},\ \bibinfo {pages} {1} (\bibinfo {year} {2016})},\
  \Eprint {http://arxiv.org/abs/1402.6703}{arXiv:1402.6703}\BibitemShut
  {NoStop}%
\bibitem [{\citenamefont {Zhou}\ \emph {{\it et~al.}}(2015)\citenamefont
  {Zhou}, \citenamefont {Liang}, \citenamefont {Huang}, \citenamefont {Li},
  \citenamefont {Fan}, \citenamefont {Feng},\ and\ \citenamefont
  {Chang}}]{Zhou:2014lva}%
  \BibitemOpen
  \bibfield  {author} {\bibinfo {author} {\bibfnamefont {B.}~\bibnamefont
  {Zhou}}, \bibinfo {author} {\bibfnamefont {Y.-F.}\ \bibnamefont {Liang}},
  \bibinfo {author} {\bibfnamefont {X.}~\bibnamefont {Huang}}, \bibinfo
  {author} {\bibfnamefont {X.}~\bibnamefont {Li}}, \bibinfo {author}
  {\bibfnamefont {Y.-Z.}\ \bibnamefont {Fan}}, \bibinfo {author} {\bibfnamefont
  {L.}~\bibnamefont {Feng}}, and\ \bibinfo {author} {\bibfnamefont
  {J.}~\bibnamefont {Chang}},\ }\bibfield  {title} {\enquote {\bibinfo {title}
  {{GeV excess in the Milky Way: The role of diffuse galactic gamma-ray
  emission templates}},}\ }\href {\doibase 10.1103/PhysRevD.91.123010}
  {\bibfield  {journal} {\bibinfo  {journal} {Phys. Rev. D}\ }\textbf {\bibinfo
  {volume} {91}},\ \bibinfo {pages} {123010} (\bibinfo {year} {2015})},\
  \Eprint {http://arxiv.org/abs/1406.6948}{arXiv:1406.6948}\BibitemShut
  {NoStop}%
\bibitem [{\citenamefont {Calore}\ \emph {{\it et~al.}}(2015)\citenamefont
  {Calore}, \citenamefont {Cholis},\ and\ \citenamefont
  {Weniger}}]{Calore:2014xka}%
  \BibitemOpen
  \bibfield  {author} {\bibinfo {author} {\bibfnamefont {F.}~\bibnamefont
  {Calore}}, \bibinfo {author} {\bibfnamefont {I.}~\bibnamefont {Cholis}}, and\
  \bibinfo {author} {\bibfnamefont {C.}~\bibnamefont {Weniger}},\ }\bibfield
  {title} {\enquote {\bibinfo {title} {{Background model systematics for the
  Fermi GeV excess}},}\ }\href {\doibase 10.1088/1475-7516/2015/03/038}
  {\bibfield  {journal} {\bibinfo  {journal} {JCAP}\ }\textbf {\bibinfo
  {volume} {1503}},\ \bibinfo {pages} {038} (\bibinfo {year} {2015})},\ \Eprint
  {http://arxiv.org/abs/1409.0042}{arXiv:1409.0042}\BibitemShut {NoStop}%
\bibitem [{\citenamefont {{Huang}}\ \emph {{\it et~al.}}(2016)\citenamefont
  {{Huang}}, \citenamefont {{En{\ss}lin}},\ and\ \citenamefont
  {{Selig}}}]{Huang:2015rlu}%
  \BibitemOpen
  \bibfield  {author} {\bibinfo {author} {\bibfnamefont {X.}~\bibnamefont
  {{Huang}}}, \bibinfo {author} {\bibfnamefont {T.}~\bibnamefont
  {{En{\ss}lin}}}, and\ \bibinfo {author} {\bibfnamefont {M.}~\bibnamefont
  {{Selig}}},\ }\bibfield  {title} {\enquote {\bibinfo {title} {{Galactic dark
  matter search via phenomenological astrophysics modeling}},}\ }\href
  {\doibase 10.1088/1475-7516/2016/04/030} {\bibfield  {journal} {\bibinfo
  {journal} {\jcap}\ }\textbf {\bibinfo {volume} {4}},\ \bibinfo {eid} {030}
  (\bibinfo {year} {2016})},\ \Eprint
  {http://arxiv.org/abs/1511.02621}{arXiv:1511.02621}\BibitemShut {NoStop}%
\bibitem [{\citenamefont {{Ackermann}}\ \emph {{\it et~al.}}(2017)\citenamefont
  {{Ackermann}}, \citenamefont {{Ajello}}, \citenamefont {{Albert}},
  \citenamefont {{Atwood}}, \citenamefont {{Baldini}}, \citenamefont
  {{Ballet}}, \citenamefont {{Barbiellini}}, \citenamefont {{Bastieri}},
  \citenamefont {{Bellazzini}}, \citenamefont {{Bissaldi}}, \citenamefont
  {{Blandford}}, \citenamefont {{Bloom}}, \citenamefont {{Bonino}},
  \citenamefont {{Bottacini}}, \citenamefont {{Brandt}}, \citenamefont
  {{Bregeon}}, \citenamefont {{Bruel}}, \citenamefont {{Buehler}},
  \citenamefont {{Burnett}}, \citenamefont {{Cameron}}, \citenamefont
  {{Caputo}}, \citenamefont {{Caragiulo}}, \citenamefont {{Caraveo}},
  \citenamefont {{Cavazzuti}}, \citenamefont {{Cecchi}}, \citenamefont
  {{Charles}}, \citenamefont {{Chekhtman}}, \citenamefont {{Chiang}},
  \citenamefont {{Chiappo}}, \citenamefont {{Chiaro}}, \citenamefont
  {{Ciprini}}, \citenamefont {{Conrad}}, \citenamefont {{Costanza}},
  \citenamefont {{Cuoco}}, \citenamefont {{Cutini}}, \citenamefont
  {{D'Ammando}}, \citenamefont {{de Palma}}, \citenamefont {{Desiante}},
  \citenamefont {{Digel}}, \citenamefont {{Di Lalla}}, \citenamefont {{Di
  Mauro}}, \citenamefont {{Di Venere}}, \citenamefont {{Drell}}, \citenamefont
  {{Favuzzi}}, \citenamefont {{Fegan}}, \citenamefont {{Ferrara}},
  \citenamefont {{Focke}}, \citenamefont {{Franckowiak}}, \citenamefont
  {{Fukazawa}}, \citenamefont {{Funk}}, \citenamefont {{Fusco}}, \citenamefont
  {{Gargano}}, \citenamefont {{Gasparrini}}, \citenamefont {{Giglietto}},
  \citenamefont {{Giordano}}, \citenamefont {{Giroletti}}, \citenamefont
  {{Glanzman}}, \citenamefont {{Gomez-Vargas}}, \citenamefont {{Green}},
  \citenamefont {{Grenier}}, \citenamefont {{Grove}}, \citenamefont
  {{Guillemot}}, \citenamefont {{Guiriec}}, \citenamefont {{Gustafsson}},
  \citenamefont {{Harding}}, \citenamefont {{Hays}}, \citenamefont {{Hewitt}},
  \citenamefont {{Horan}}, \citenamefont {{Jogler}}, \citenamefont {{Johnson}},
  \citenamefont {{Kamae}}, \citenamefont {{Kocevski}}, \citenamefont {{Kuss}},
  \citenamefont {{La Mura}}, \citenamefont {{Larsson}}, \citenamefont
  {{Latronico}}, \citenamefont {{Li}}, \citenamefont {{Longo}}, \citenamefont
  {{Loparco}}, \citenamefont {{Lovellette}}, \citenamefont {{Lubrano}},
  \citenamefont {{Magill}}, \citenamefont {{Maldera}}, \citenamefont
  {{Malyshev}}, \citenamefont {{Manfreda}}, \citenamefont {{Martin}},
  \citenamefont {{Mazziotta}}, \citenamefont {{Michelson}}, \citenamefont
  {{Mirabal}}, \citenamefont {{Mitthumsiri}}, \citenamefont {{Mizuno}},
  \citenamefont {{Moiseev}}, \citenamefont {{Monzani}}, \citenamefont
  {{Morselli}}, \citenamefont {{Negro}}, \citenamefont {{Nuss}}, \citenamefont
  {{Ohsugi}}, \citenamefont {{Orienti}}, \citenamefont {{Orlando}},
  \citenamefont {{Ormes}}, \citenamefont {{Paneque}}, \citenamefont
  {{Perkins}}, \citenamefont {{Persic}}, \citenamefont {{Pesce-Rollins}},
  \citenamefont {{Piron}}, \citenamefont {{Principe}}, \citenamefont
  {{Rain{\`o}}}, \citenamefont {{Rando}}, \citenamefont {{Razzano}},
  \citenamefont {{Razzaque}}, \citenamefont {{Reimer}}, \citenamefont
  {{Reimer}}, \citenamefont {{S{\'a}nchez-Conde}}, \citenamefont {{Sgr{\`o}}},
  \citenamefont {{Simone}}, \citenamefont {{Siskind}}, \citenamefont {{Spada}},
  \citenamefont {{Spandre}}, \citenamefont {{Spinelli}}, \citenamefont
  {{Suson}}, \citenamefont {{Tajima}}, \citenamefont {{Tanaka}}, \citenamefont
  {{Thayer}}, \citenamefont {{Tibaldo}}, \citenamefont {{Torres}},
  \citenamefont {{Troja}}, \citenamefont {{Uchiyama}}, \citenamefont
  {{Vianello}}, \citenamefont {{Wood}}, \citenamefont {{Wood}}, \citenamefont
  {{Zaharijas}}, \citenamefont {{Zimmer}},\ and\ \citenamefont {{Fermi LAT
  Collaboration}}}]{fermi17GCE}%
  \BibitemOpen
  \bibfield  {author} {\bibinfo {author} {\bibfnamefont {M.}~\bibnamefont
  {{Ackermann}}} {\it et~al.},\ }\bibfield  {title} {\enquote {\bibinfo {title}
  {{The Fermi Galactic Center GeV Excess and Implications for Dark Matter}},}\
  }\href {\doibase 10.3847/1538-4357/aa6cab} {\bibfield  {journal} {\bibinfo
  {journal} {Astrophys. J.}\ }\textbf {\bibinfo {volume} {840}},\ \bibinfo
  {eid} {43} (\bibinfo {year} {2017})},\ \Eprint
  {http://arxiv.org/abs/1704.03910}{arXiv:1704.03910}\BibitemShut {NoStop}%
\bibitem [{\citenamefont {Abdughani}\ \emph {{\it et~al.}}(2021)\citenamefont
  {Abdughani}, \citenamefont {Fan}, \citenamefont {Feng}, \citenamefont {Tsai},
  \citenamefont {Wu},\ and\ \citenamefont {Yuan}}]{ABDUGHANI2021}%
  \BibitemOpen
  \bibfield  {author} {\bibinfo {author} {\bibfnamefont {M.}~\bibnamefont
  {Abdughani}}, \bibinfo {author} {\bibfnamefont {Y.-Z.}\ \bibnamefont {Fan}},
  \bibinfo {author} {\bibfnamefont {L.}~\bibnamefont {Feng}}, \bibinfo {author}
  {\bibfnamefont {Y.-L.~S.}\ \bibnamefont {Tsai}}, \bibinfo {author}
  {\bibfnamefont {L.}~\bibnamefont {Wu}}, and\ \bibinfo {author} {\bibfnamefont
  {Q.}~\bibnamefont {Yuan}},\ }\bibfield  {title} {\enquote {\bibinfo {title}
  {A common origin of muon g-2 anomaly, galaxy center gev excess and ams-02
  anti-proton excess in the nmssm},}\ }\href {\doibase
  https://doi.org/10.1016/j.scib.2021.07.029} {\bibfield  {journal} {\bibinfo
  {journal} {Science Bulletin}\ } (\bibinfo {year} {2021}),\
  https://doi.org/10.1016/j.scib.2021.07.029}\BibitemShut {NoStop}%
\bibitem [{\citenamefont {{Mattox}}\ \emph {{\it et~al.}}(1996)\citenamefont
  {{Mattox}}, \citenamefont {{Bertsch}}, \citenamefont {{Chiang}},
  \citenamefont {{Dingus}}, \citenamefont {{Digel}}, \citenamefont
  {{Esposito}}, \citenamefont {{Fierro}}, \citenamefont {{Hartman}},
  \citenamefont {{Hunter}}, \citenamefont {{Kanbach}}, \citenamefont
  {{Kniffen}}, \citenamefont {{Lin}}, \citenamefont {{Macomb}}, \citenamefont
  {{Mayer-Hasselwander}}, \citenamefont {{Michelson}}, \citenamefont {{von
  Montigny}}, \citenamefont {{Mukherjee}}, \citenamefont {{Nolan}},
  \citenamefont {{Ramanamurthy}}, \citenamefont {{Schneid}}, \citenamefont
  {{Sreekumar}}, \citenamefont {{Thompson}},\ and\ \citenamefont
  {{Willis}}}]{1996ApJ...461..396M}%
  \BibitemOpen
  \bibfield  {author} {\bibinfo {author} {\bibfnamefont {J.~R.}\ \bibnamefont
  {{Mattox}}} {\it et~al.},\ }\bibfield  {title} {\enquote {\bibinfo {title}
  {{The Likelihood Analysis of EGRET Data}},}\ }\href {\doibase 10.1086/177068}
  {\bibfield  {journal} {\bibinfo  {journal} {Astrophys. J.}\ }\textbf
  {\bibinfo {volume} {461}},\ \bibinfo {pages} {396} (\bibinfo {year}
  {1996})}\BibitemShut {NoStop}%
\bibitem [{\citenamefont {Cirelli}\ \emph {{\it et~al.}}(2011)\citenamefont
  {Cirelli}, \citenamefont {Corcella}, \citenamefont {Hektor}, \citenamefont
  {Hutsi}, \citenamefont {Kadastik}, \citenamefont {Panci}, \citenamefont
  {Raidal}, \citenamefont {Sala},\ and\ \citenamefont
  {Strumia}}]{Cirelli:2010xx}%
  \BibitemOpen
  \bibfield  {author} {\bibinfo {author} {\bibfnamefont {M.}~\bibnamefont
  {Cirelli}}, \bibinfo {author} {\bibfnamefont {G.}~\bibnamefont {Corcella}},
  \bibinfo {author} {\bibfnamefont {A.}~\bibnamefont {Hektor}}, \bibinfo
  {author} {\bibfnamefont {G.}~\bibnamefont {Hutsi}}, \bibinfo {author}
  {\bibfnamefont {M.}~\bibnamefont {Kadastik}}, \bibinfo {author}
  {\bibfnamefont {P.}~\bibnamefont {Panci}}, \bibinfo {author} {\bibfnamefont
  {M.}~\bibnamefont {Raidal}}, \bibinfo {author} {\bibfnamefont
  {F.}~\bibnamefont {Sala}}, and\ \bibinfo {author} {\bibfnamefont
  {A.}~\bibnamefont {Strumia}},\ }\bibfield  {title} {\enquote {\bibinfo
  {title} {{PPPC 4 DM ID: A Poor Particle Physicist Cookbook for Dark Matter
  Indirect Detection}},}\ }\href {\doibase 10.1088/1475-7516/2012/10/E01,
  10.1088/1475-7516/2011/03/051} {\bibfield  {journal} {\bibinfo  {journal}
  {JCAP}\ }\textbf {\bibinfo {volume} {1103}},\ \bibinfo {pages} {051}
  (\bibinfo {year} {2011})},\ \bibinfo {note} {[Erratum: JCAP1210,E01(2012)]},\
  \Eprint {http://arxiv.org/abs/1012.4515}{arXiv:1012.4515}\BibitemShut
  {NoStop}%
\bibitem [{\citenamefont {Evans}\ \emph {{\it et~al.}}(2016)\citenamefont
  {Evans}, \citenamefont {Sanders},\ and\ \citenamefont
  {Geringer-Sameth}}]{PhysRevD.93.103512}%
  \BibitemOpen
  \bibfield  {author} {\bibinfo {author} {\bibfnamefont {N.~W.}\ \bibnamefont
  {Evans}}, \bibinfo {author} {\bibfnamefont {J.~L.}\ \bibnamefont {Sanders}},
  and\ \bibinfo {author} {\bibfnamefont {A.}~\bibnamefont {Geringer-Sameth}},\
  }\bibfield  {title} {\enquote {\bibinfo {title} {Simple j-factors and
  d-factors for indirect dark matter detection},}\ }\href {\doibase
  10.1103/PhysRevD.93.103512} {\bibfield  {journal} {\bibinfo  {journal} {Phys.
  Rev. D}\ }\textbf {\bibinfo {volume} {93}},\ \bibinfo {pages} {103512}
  (\bibinfo {year} {2016})}\BibitemShut {NoStop}%
\bibitem [{\citenamefont {Sanders}\ \emph {{\it et~al.}}(2016)\citenamefont
  {Sanders}, \citenamefont {Evans}, \citenamefont {Geringer-Sameth},\ and\
  \citenamefont {Dehnen}}]{PhysRevD.94.063521}%
  \BibitemOpen
  \bibfield  {author} {\bibinfo {author} {\bibfnamefont {J.~L.}\ \bibnamefont
  {Sanders}}, \bibinfo {author} {\bibfnamefont {N.~W.}\ \bibnamefont {Evans}},
  \bibinfo {author} {\bibfnamefont {A.}~\bibnamefont {Geringer-Sameth}}, and\
  \bibinfo {author} {\bibfnamefont {W.}~\bibnamefont {Dehnen}},\ }\bibfield
  {title} {\enquote {\bibinfo {title} {Indirect dark matter detection for
  flattened dwarf galaxies},}\ }\href {\doibase 10.1103/PhysRevD.94.063521}
  {\bibfield  {journal} {\bibinfo  {journal} {Phys. Rev. D}\ }\textbf {\bibinfo
  {volume} {94}},\ \bibinfo {pages} {063521} (\bibinfo {year}
  {2016})}\BibitemShut {NoStop}%
\bibitem [{\citenamefont {Navarro}\ \emph {{\it et~al.}}(1997)\citenamefont
  {Navarro}, \citenamefont {Frenk},\ and\ \citenamefont
  {White}}]{Navarro:1996gj}%
  \BibitemOpen
  \bibfield  {author} {\bibinfo {author} {\bibfnamefont {J.~F.}\ \bibnamefont
  {Navarro}}, \bibinfo {author} {\bibfnamefont {C.~S.}\ \bibnamefont {Frenk}},
  and\ \bibinfo {author} {\bibfnamefont {S.~D.~M.}\ \bibnamefont {White}},\
  }\bibfield  {title} {\enquote {\bibinfo {title} {{A Universal density profile
  from hierarchical clustering}},}\ }\href {\doibase 10.1086/304888} {\bibfield
   {journal} {\bibinfo  {journal} {Astrophys. J.}\ }\textbf {\bibinfo {volume}
  {490}},\ \bibinfo {pages} {493} (\bibinfo {year} {1997})},\ \Eprint
  {http://arxiv.org/abs/astro-ph/9611107}{arXiv:astro-ph/9611107}\BibitemShut
  {NoStop}%
\bibitem [{\citenamefont {Willman}\ \emph {{\it et~al.}}(2011)\citenamefont
  {Willman}, \citenamefont {Geha}, \citenamefont {Strader}, \citenamefont
  {Strigari}, \citenamefont {Simon}, \citenamefont {Kirby}, \citenamefont
  {Ho},\ and\ \citenamefont {Warres}}]{Willman_2011}%
  \BibitemOpen
  \bibfield  {author} {\bibinfo {author} {\bibfnamefont {B.}~\bibnamefont
  {Willman}}, \bibinfo {author} {\bibfnamefont {M.}~\bibnamefont {Geha}},
  \bibinfo {author} {\bibfnamefont {J.}~\bibnamefont {Strader}}, \bibinfo
  {author} {\bibfnamefont {L.~E.}\ \bibnamefont {Strigari}}, \bibinfo {author}
  {\bibfnamefont {J.~D.}\ \bibnamefont {Simon}}, \bibinfo {author}
  {\bibfnamefont {E.}~\bibnamefont {Kirby}}, \bibinfo {author} {\bibfnamefont
  {N.}~\bibnamefont {Ho}}, and\ \bibinfo {author} {\bibfnamefont
  {A.}~\bibnamefont {Warres}},\ }\bibfield  {title} {\enquote {\bibinfo {title}
  {{WILLMAN} 1{\textemdash}a {PROBABLE} {DWARF} {GALAXY} {WITH} {AN}
  {IRREGULAR} {KINEMATIC} {DISTRIBUTION}},}\ }\href {\doibase
  10.1088/0004-6256/142/4/128} {\bibfield  {journal} {\bibinfo  {journal} {The
  Astronomical Journal}\ }\textbf {\bibinfo {volume} {142}},\ \bibinfo {pages}
  {128} (\bibinfo {year} {2011})}\BibitemShut {NoStop}%
\bibitem [{\citenamefont {{Massaro}}\ \emph {{\it et~al.}}(2009)\citenamefont
  {{Massaro}}, \citenamefont {{Giommi}}, \citenamefont {{Leto}}, \citenamefont
  {{Marchegiani}}, \citenamefont {{Maselli}}, \citenamefont {{Perri}},
  \citenamefont {{Piranomonte}},\ and\ \citenamefont
  {{Sclavi}}}]{2009A&A...495..691M}%
  \BibitemOpen
  \bibfield  {author} {\bibinfo {author} {\bibfnamefont {E.}~\bibnamefont
  {{Massaro}}}, \bibinfo {author} {\bibfnamefont {P.}~\bibnamefont {{Giommi}}},
  \bibinfo {author} {\bibfnamefont {C.}~\bibnamefont {{Leto}}}, \bibinfo
  {author} {\bibfnamefont {P.}~\bibnamefont {{Marchegiani}}}, \bibinfo {author}
  {\bibfnamefont {A.}~\bibnamefont {{Maselli}}}, \bibinfo {author}
  {\bibfnamefont {M.}~\bibnamefont {{Perri}}}, \bibinfo {author} {\bibfnamefont
  {S.}~\bibnamefont {{Piranomonte}}}, and\ \bibinfo {author} {\bibfnamefont
  {S.}~\bibnamefont {{Sclavi}}},\ }\bibfield  {title} {\enquote {\bibinfo
  {title} {{Roma-BZCAT: a multifrequency catalogue of blazars}},}\ }\href
  {\doibase 10.1051/0004-6361:200810161} {\bibfield  {journal} {\bibinfo
  {journal} {\aap}\ }\textbf {\bibinfo {volume} {495}},\ \bibinfo {pages} {691}
  (\bibinfo {year} {2009})},\ \Eprint
  {http://arxiv.org/abs/0810.2206}{arXiv:0810.2206}\BibitemShut {NoStop}%
\bibitem [{\citenamefont {{Healey}}\ \emph {{\it et~al.}}(2007)\citenamefont
  {{Healey}}, \citenamefont {{Romani}}, \citenamefont {{Taylor}}, \citenamefont
  {{Sadler}}, \citenamefont {{Ricci}}, \citenamefont {{Murphy}}, \citenamefont
  {{Ulvestad}},\ and\ \citenamefont {{Winn}}}]{2007ApJS..171...61H}%
  \BibitemOpen
  \bibfield  {author} {\bibinfo {author} {\bibfnamefont {S.~E.}\ \bibnamefont
  {{Healey}}}, \bibinfo {author} {\bibfnamefont {R.~W.}\ \bibnamefont
  {{Romani}}}, \bibinfo {author} {\bibfnamefont {G.~B.}\ \bibnamefont
  {{Taylor}}}, \bibinfo {author} {\bibfnamefont {E.~M.}\ \bibnamefont
  {{Sadler}}}, \bibinfo {author} {\bibfnamefont {R.}~\bibnamefont {{Ricci}}},
  \bibinfo {author} {\bibfnamefont {T.}~\bibnamefont {{Murphy}}}, \bibinfo
  {author} {\bibfnamefont {J.~S.}\ \bibnamefont {{Ulvestad}}}, and\ \bibinfo
  {author} {\bibfnamefont {J.~N.}\ \bibnamefont {{Winn}}},\ }\bibfield  {title}
  {\enquote {\bibinfo {title} {{CRATES: An All-Sky Survey of Flat-Spectrum
  Radio Sources}},}\ }\href {\doibase 10.1086/513742} {\bibfield  {journal}
  {\bibinfo  {journal} {Astrophys. J. Suppl.}\ }\textbf {\bibinfo {volume}
  {171}},\ \bibinfo {pages} {61} (\bibinfo {year} {2007})},\ \Eprint
  {http://arxiv.org/abs/astro-ph/0702346}{arXiv:astro-ph/0702346}\BibitemShut
  {NoStop}%
\bibitem [{\citenamefont {{Healey}}\ \emph {{\it et~al.}}(2008)\citenamefont
  {{Healey}}, \citenamefont {{Romani}}, \citenamefont {{Cotter}}, \citenamefont
  {{Michelson}}, \citenamefont {{Schlafly}}, \citenamefont {{Readhead}},
  \citenamefont {{Giommi}}, \citenamefont {{Chaty}}, \citenamefont
  {{Grenier}},\ and\ \citenamefont {{Weintraub}}}]{2008ApJS..175...97H}%
  \BibitemOpen
  \bibfield  {author} {\bibinfo {author} {\bibfnamefont {S.~E.}\ \bibnamefont
  {{Healey}}}, \bibinfo {author} {\bibfnamefont {R.~W.}\ \bibnamefont
  {{Romani}}}, \bibinfo {author} {\bibfnamefont {G.}~\bibnamefont {{Cotter}}},
  \bibinfo {author} {\bibfnamefont {P.~F.}\ \bibnamefont {{Michelson}}},
  \bibinfo {author} {\bibfnamefont {E.~F.}\ \bibnamefont {{Schlafly}}},
  \bibinfo {author} {\bibfnamefont {A.~C.~S.}\ \bibnamefont {{Readhead}}},
  \bibinfo {author} {\bibfnamefont {P.}~\bibnamefont {{Giommi}}}, \bibinfo
  {author} {\bibfnamefont {S.}~\bibnamefont {{Chaty}}}, \bibinfo {author}
  {\bibfnamefont {I.~A.}\ \bibnamefont {{Grenier}}}, and\ \bibinfo {author}
  {\bibfnamefont {L.~C.}\ \bibnamefont {{Weintraub}}},\ }\bibfield  {title}
  {\enquote {\bibinfo {title} {{CGRaBS: An All-Sky Survey of Gamma-Ray Blazar
  Candidates}},}\ }\href {\doibase 10.1086/523302} {\bibfield  {journal}
  {\bibinfo  {journal} {Astrophys. J. Suppl.}\ }\textbf {\bibinfo {volume}
  {175}},\ \bibinfo {pages} {97} (\bibinfo {year} {2008})},\ \Eprint
  {http://arxiv.org/abs/0709.1735}{arXiv:0709.1735}\BibitemShut {NoStop}%
\bibitem [{\citenamefont {{D'Abrusco}}\ \emph {{\it et~al.}}(2014)\citenamefont
  {{D'Abrusco}}, \citenamefont {{Massaro}}, \citenamefont {{Paggi}},
  \citenamefont {{Smith}}, \citenamefont {{Masetti}}, \citenamefont
  {{Landoni}},\ and\ \citenamefont {{Tosti}}}]{2014ApJS..215...14D}%
  \BibitemOpen
  \bibfield  {author} {\bibinfo {author} {\bibfnamefont {R.}~\bibnamefont
  {{D'Abrusco}}}, \bibinfo {author} {\bibfnamefont {F.}~\bibnamefont
  {{Massaro}}}, \bibinfo {author} {\bibfnamefont {A.}~\bibnamefont {{Paggi}}},
  \bibinfo {author} {\bibfnamefont {H.~A.}\ \bibnamefont {{Smith}}}, \bibinfo
  {author} {\bibfnamefont {N.}~\bibnamefont {{Masetti}}}, \bibinfo {author}
  {\bibfnamefont {M.}~\bibnamefont {{Landoni}}}, and\ \bibinfo {author}
  {\bibfnamefont {G.}~\bibnamefont {{Tosti}}},\ }\bibfield  {title} {\enquote
  {\bibinfo {title} {{The WISE Blazar-like Radio-loud Sources: An All-sky
  Catalog of Candidate {\ensuremath{\gamma}}-ray Blazars}},}\ }\href {\doibase
  10.1088/0067-0049/215/1/14} {\bibfield  {journal} {\bibinfo  {journal}
  {Astrophys. J. Suppl.}\ }\textbf {\bibinfo {volume} {215}},\ \bibinfo {eid}
  {14} (\bibinfo {year} {2014})},\ \Eprint
  {http://arxiv.org/abs/1410.0029}{arXiv:1410.0029}\BibitemShut {NoStop}%
\bibitem [{\citenamefont {{Galper}}\ \emph {{\it et~al.}}(2013)\citenamefont
  {{Galper}}, \citenamefont {{Adriani}}, \citenamefont {{Aptekar}},
  \citenamefont {{Arkhangelskaja}}, \citenamefont {{Arkhangelskiy}},
  \citenamefont {{Boezio}}, \citenamefont {{Bonvicini}}, \citenamefont
  {{Boyarchuk}}, \citenamefont {{Gusakov}}, \citenamefont {{Farber}},
  \citenamefont {{Fradkin}}, \citenamefont {{Kachanov}}, \citenamefont
  {{Kaplin}}, \citenamefont {{Kheymits}}, \citenamefont {{Leonov}},
  \citenamefont {{Longo}}, \citenamefont {{Maestro}}, \citenamefont
  {{Marrocchesi}}, \citenamefont {{Mazets}}, \citenamefont {{Mocchiutti}},
  \citenamefont {{Moiseev}}, \citenamefont {{Mori}}, \citenamefont
  {{Moskalenko}}, \citenamefont {{Naumov}}, \citenamefont {{Papini}},
  \citenamefont {{Picozza}}, \citenamefont {{Rodin}}, \citenamefont {{Runtso}},
  \citenamefont {{Sparvoli}}, \citenamefont {{Spillantini}}, \citenamefont
  {{Suchkov}}, \citenamefont {{Tavani}}, \citenamefont {{Topchiev}},
  \citenamefont {{Vacchi}}, \citenamefont {{Vannuccini}}, \citenamefont
  {{Yurkin}}, \citenamefont {{Zampa}},\ and\ \citenamefont
  {{Zverev}}}]{gamma400}%
  \BibitemOpen
  \bibfield  {author} {\bibinfo {author} {\bibfnamefont {A.~M.}\ \bibnamefont
  {{Galper}}} {\it et~al.},\ }\bibfield  {title} {\enquote {\bibinfo {title}
  {{Status of the GAMMA-400 project}},}\ }\href {\doibase
  10.1016/j.asr.2012.01.019} {\bibfield  {journal} {\bibinfo  {journal}
  {Advances in Space Research}\ }\textbf {\bibinfo {volume} {51}},\ \bibinfo
  {pages} {297} (\bibinfo {year} {2013})},\ \Eprint
  {http://arxiv.org/abs/1201.2490}{arXiv:1201.2490}\BibitemShut {NoStop}%
\bibitem [{\citenamefont {{Zhang}}\ \emph {{\it et~al.}}(2014)\citenamefont
  {{Zhang}} {\it et~al.}}]{zhang14herd}%
  \BibitemOpen
  \bibfield  {author} {\bibinfo {author} {\bibfnamefont {S.~N.}\ \bibnamefont
  {{Zhang}}}  {\it et~al.} (\bibinfo {collaboration} {HERD} Collaboration),\
  }\bibfield  {title} {\enquote {\bibinfo {title} {{The high energy
  cosmic-radiation detection (HERD) facility onboard China's Space Station}},}\
  }\href {\doibase 10.1117/12.2055280} {\bibfield  {journal} {\bibinfo
  {journal} {Proc. SPIE Int. Soc. Opt. Eng.}\ }\textbf {\bibinfo {volume}
  {9144}},\ \bibinfo {eid} {91440X} (\bibinfo {year} {2014})},\ \Eprint
  {http://arxiv.org/abs/1407.4866}{arXiv:1407.4866}\BibitemShut {NoStop}%
\end{thebibliography}
 %
                                                                                    
\end{document}